\documentclass[letterpaper,12pt]{article}
\usepackage[english]{babel}
\usepackage[utf8]{inputenc}
\usepackage[left=2.5cm,right=2.5cm,top=2.5cm,bottom=2.5cm]{geometry}
\usepackage{enumerate}
\usepackage{graphicx}
\usepackage{amssymb,amsfonts,amsmath,bm}
\usepackage{amsthm}
\usepackage{epstopdf}
\usepackage{latexsym}
\usepackage{color}
\usepackage{comment}

\theoremstyle{plain} 
\newtheorem{theorem}{Theorem}

\newtheorem{Assumption}{Assumption}
\newtheorem{Lemma}{Lemma}

\def\sym#1{\ifmmode^{#1}\else\(^{#1}\)\fi}

\newcommand{\cvd}{\mbox{$\stackrel{d}{\longrightarrow}\,$}}

\newcommand\numberthis{\addtocounter{equation}{1}\tag{\theequation}}
\newcommand{\pl}{\mbox{\rm plim}}
\newcommand{\Var}{\mbox{\rm Var}}

\def\esp{\mathbb{E}}
\def\ind{\mathbb{I}}

\DeclareMathOperator*{\argmax}{arg\,max}

\usepackage{multirow}
\usepackage{tabularx}
\newcolumntype{C}{>{\centering\arraybackslash}X} 
\newcolumntype{Y}{>{\centering\arraybackslash}X}
\newcolumntype{H}{>{\setbox0=\hbox\bgroup}c<{\egroup}@{}}
\newcolumntype{P}[1]{>{\centering\arraybackslash}p{#1}}

\usepackage[title]{appendix} 
\usepackage{longtable}
\usepackage{pdflscape}
\usepackage{booktabs}

\usepackage[flushleft]{threeparttable}
\usepackage{threeparttablex} 

\usepackage[font=bf, justification=centering, width = \textwidth]{caption}
\usepackage{subcaption}
\usepackage{array}

\newcounter{mysubtable}

\usepackage[nodisplayskipstretch]{setspace}
\usepackage{footnote}
\usepackage{har2nat} 
\usepackage{hyperref}
\hypersetup{
	colorlinks=true,
	linkcolor = black,
	citecolor = blue,
	urlcolor=black,
	pdfborder={0 0 0},
}

\makeatletter
\def\@xfootnote[#1]{%
	\protected@xdef\@thefnmark{#1}%
	\@footnotemark\@footnotetext}
\makeatother

\newlength{\saveparindent}
\begin{document}


\begin{titlepage}
\title{\Large \textbf{Estimating Program Participation with Partial Validation%
      \footnote{We thank Samuel Asare, Chris Bollinger, Desire Kedagni and conference participants at the Southern Economic Association meeting for helpful comments.}}
    \vspace{1.5cm}}
  \author{%
    Augustine Denteh%
    \footnote{Department of Economics, Davidson College,
			209 Ridge Rd, Box 5000, Davidson, NC 28035;  
			Email: \href{mailto:audenteh@davidson.edu.edu}{audenteh@davidson.edu}; Telephone: 704-894-2504.}
	\and
    Pierre E. Nguimkeu%
    \footnote{Correspondence: Department of Economics, Andrew Young School of Policy Studies, Georgia State University, P.O. Box 3992, Atlanta, GA 30302-3992;
      \href{mailto:nnguimkeu@gsu.edu}{nnguimkeu@gsu.edu}.}
  }%
  \date{December 2025}
  \maketitle

\begin{singlespace}
    \begin{abstract}
      \noindent
This paper considers the estimation of binary choice models when survey responses are possibly misclassified but one of the response category can be validated. Partial validation may occur when survey questions about participation include follow-up questions on that particular response category. In this case, we show that the initial two-sided misclassification problem can be transformed into a one-sided one, based on the partially validated responses. Using the updated responses naively for estimation does not solve or mitigate the misclassification bias, and we derive the ensuing asymptotic bias under general conditions. We then show how the partially validated responses can be used to construct a model for participation and propose consistent and asymptotically normal estimators that overcome misclassification error. Monte Carlo simulations are provided to demonstrate the finite sample performance of the proposed and selected existing methods. We provide an empirical illustration on the determinants of health insurance coverage in Ghana. We discuss implications for the design of survey questionnaires that allow researchers to overcome misclassification biases without recourse to relatively costly and often imperfect validation data. 

\vspace{0.5cm}
\end{abstract}
\end{singlespace}

\end{titlepage}

\onehalfspacing
\pagenumbering{arabic} 

\section{Introduction}\label{intro}
Many important questions in economics concern estimating regression models for binary outcome variables. Such binary outcomes are frequently subject to misclassification errors. Several papers have documented the rising rates of misreporting in binary measures such as participation in government programs \citep{MarquisMoore1990,meyer2019misreporting, meyer2021use,celhay2021errors, meyer2022errors,celhay2024leads,courtemanche2019estimating}, education \citep{kane1995varied,black2003measurement}, labor union membership \citep{freeman1984longitudinal,card1996effect}, employment status \citep{bound1991extent}, voting behavior \citep{ansolabehere2012validation} and health insurance \cite{call2008medicaid,davern2009partially}. Regarding participation in social programs, misreporting rates (false negatives) can sometimes exceed 50\%. Two excellent surveys of the prevalence of measurement error in various contexts are \citet{bound2001measurement} and \citet{meyer2015household}. Due to the inherently non-classical nature of measurement errors in binary variables, the coefficient estimates from such binary choice models may be substantially biased. In most instances, access to validation data with a measure of ``truth" may be difficult or impossible. The typical remedy for measurement error is to resort to econometric modeling approaches, often with convenient assumptions on how misclassification probabilities depend on covariates.\\

In this paper, we study the identification and estimation of binary choice models when the dependent variable is subject to covariate-dependent and endogenous misclassification that can be partially validated. Partial validation occurs when the features of the survey design allow researchers to verify the accuracy of the responses for one category of the binary outcome. For example, in the 1996 Medical Expenditure Panel Survey (MEPS), health insurance status is partially validated for some respondents based on insurance cards, insurance policy documents, and interviews with employers or insurance companies \citep{hill2007accuracy,kreider2009partially}. The 1996 MEPS included a comprehensive set of health insurance questions, including questions on all jobs held by the respondent. Importantly, for each reported source of health insurance, the MEPS interviewer asked to see the respondent's insurance card and insurance policy documents. \citet{hill2007accuracy} finds that about 77\% of those who reported having private insurance coverage in the 1996 MEPS were able to produce an insurance card during the interview. 

The Ghana Standard Demographic Health Survey (Ghana-DHS) provides another example of partial validation of health insurance status, which we later use for our empirical illustration. Specifically, the Ghana-DHS respondents were asked about their participation in the National Health Insurance Scheme (NHIS), which is Ghana's flagship program introduced in 2003. Similar to the 1996 MEPS, those respondents who reported NHIS participation were asked to provide their NHIS identity cards \citep{asare2020health}.\footnote{Since it's implementation, the NHIS has attracted the attention of researchers and policymakers. See \citet{abrokwah2019social} and the references therein for studies examining the NHIS; for specific papers that use the Ghana-DHS to study the NHIS, see \citeasnoun{dixon2011ghana,brugiavini2016extending,ameyaw2017national}, etc.}\\

Similar scenarios occur in other contexts beyond health insurance. For instance, informality status is often misreported in survey data in developing countries. In the Cameroonian Survey on Employment and the Informal Sector (EESI), respondents are asked about the formality status of their businesses (formal or informal). Again, those who responded that their businesses were ``formal'' were asked to provide their tax registration certificate as proof of formality status while those who stated that their businesses were ``informal'' were not asked any follow-up questions. In the above examples, the validation was partial because those who did not report having any insurance or having formal businesses  were not asked to produce any documentary proof. The above validation does not provide ``gold standard" evidence of truth because the inability to show the insurance card to the interviewer is not conclusive evidence of the lack of insurance coverage. However, such limitations may apply to standard administrative sources of validation data \cite{courtemanche2019estimating}. Subject to this caveat, we leverage partial validation to create a structure that motivates consistent estimation of binary choice models with a misclassified dependent variable. \\

This paper makes three salient contributions. First, we formalize the concept of partial validation and show how it transforms two-sided misclassification into one-sided misclassification. Importantly, while this transformation does not eliminate or mitigate bias, it provides a useful framework for constructing consistent estimators. Previous studies have leveraged validation data to document the extent of misclassification and sometimes to overcome its consequences. Such efforts usually involve using  administrative data as a validation sample to obtain individual and/or aggregate misclassification rates. Examples of such studies include those that perform one-sided (partial) validation for individuals reporting participation \cite{livingston1969evaluation,marquis1978inferring,moore1996sipp}  and two-sided or complete validation \cite{MarquisMoore1990,bollinger1997modeling,bollinger2001estimation, meyer2022errors}. Most of these papers link survey responses to external validation data. Here, we rely on survey-based partial validation to update or revise the original binary measure, which is then used to estimate the binary choice model in the presence of the resulting one-sided misclassification problem. \\

Second, we derive theoretical expressions for the biases that result from using either the originally reported measure or the partially validated measure in standard binary choice models. We show that partial verification alone does not mitigate the severity of misclassification bias relative to using the originally reported measure. Using the partially validated measure in standard binary choice estimation can theoretically lead to sign reversals in estimated effects. \citet{meyer2017misclassification} also analytically derives the bias in binary choice models based on the linear probability model and the Probit model. Their results are based on directly using the originally misclassified (self-reported) participation for estimation. Here, for a general specification of a non-linear binary choice model, including the Probit model, we present new results for the bias that comes from using the partially validated participation variable as a surrogate for the true participation. Deriving the biases in both instances---using the originally reported measure or the partially validated measure---allows us to directly compare these biases and provide new insights of the consequences of two-sided and one-sided misclassification. \\

Third, we propose Maximum Likelihood Estimators (MLEs) under partial observability of the true outcome that exploits the structure created by partial validation to consistently estimate the binary choice model parameters. The existing literature commonly assumes that the probability of misclassification is constant for all individuals after we condition on observed covariates (conditionally random misclassification). Several validation studies provide empirical evidence that the conditionally random misclassification assumption can fail in practice \citep{duncan1985investigation, bound1991extent, bound1994evidence,bollinger1998measurement,black2003measurement,bollinger1997modeling,kreidersnap2012,meyer2019misreporting,meyer2017misclassification,celhay2024leads}.  In addition, the true participation decision and the misclassification processes may be arbitrarily correlated. In this paper, our framework allows for the above instances of endogenous misclassification or measurement error. Our identification strategy builds on insights from partial observability models \citep{poirier80,meng1985cost}. We establish the consistency and asymptotic normality of our proposed estimators and assess their finite sample performance through Monte Carlo simulations. We also consider an extension to semiparametric estimation. Beyond our methodological contributions, this paper has important implications for survey design. Our results suggest that incorporating verification questions for even one response category can enable researchers to obtain consistent estimates in the presence of endogenous misclassification. \\

Our work relates to a growing literature that proposes various methods for addressing misclassification in binary choice models; see \citet{meyer2017misclassification} for a recent review. \citet{meyer2017misclassification} examines the performance of six existing methods and concludes that ``applying an estimator that assumes misclassification to be conditionally random can make estimates substantively worse when this assumption is false." Such findings motivate approaches that leverage a combination of validation data and misclassification models that allow for endogenous measurement error. In this regard, much less is known about addressing bias when misclassification errors are conditionally correlated with both the true response and covariates. \citet{bollinger1997modeling} is the closest approach to using validation data to estimate a model of Food Stamp participation. In their context, \citet{bollinger1997modeling} first estimate a misclassification model (model of response error) in the validation sample drawn from the 1984 Survey of Income and Program Participation (SIPP), whose records were linked to administrative records. The estimates from the validation sample's misclassification model are then used to predict probabilities of false negatives and false positives in the primary sample of interest and maximize a pseudo-maximum likelihood to get consistent estimates of the Food Stamp participation model. \citet{meyer2017misclassification} refers to this procedure as the ``predicted probabilities estimator" and finds it to perform well. However, the predicted probabilities estimator of  \citet{bollinger1997modeling} is not feasible in our context of partial validation. As such, our paper provides a novel way to use survey-based partial validation to pursue consistent estimation of binary choice models. \\

Our simulation results demonstrate that endogenous misreporting creates severe estimation challenges for most conventional approaches, including approaches such as \citet{hausman1998misclassification}. \textcolor{black}{In contrast, the approaches based on partial observability models that we propose, including a \textit{Partial Partial Observability} MLE (PPO MLE) that we derive in this paper and the \textit{Partial Observability} MLE (PO MLE) that we borrow from the first-stage of \citet{nguimkeu2019estimation}'s estimator, perform well with the former being asymptotically more efficient than the latter under certain circumstances.} Our empirical analysis of NHIS participation in Ghana also reveals substantial differences across estimation approaches, providing strong empirical evidence for the consequences of misclassification. While naive estimators using self-reported or partially validated data show that education, religion, and ethnicity are positively associated with NHIS participation, \textcolor{black}{both our proposed estimators---PO MLE and PPO MLE---reveal opposite relationships. Comparing the two consistent approaches, the PPO MLE estimates tends to have greater precision than the PO MLE method. These findings suggest that misclassification bias can lead to misleading conclusions about the determinants of program participation.} \\

The rest of the paper is organized as follows. Section \ref{framework} presents the model framework and shows how partial verification transforms two-sided misclassification into a one-sided problem. Section \ref{biases} provides formal results on the bias due to misclassification under various scenarios. Section \ref{proposed-MLE} develops our proposed estimator and establishes its asymptotic property. Section \ref{simulations} presents Monte Carlo evidence on finite sample performance. Section \ref{empirics} contains an empirical application. Section \ref{semi-parametric} considers an extension to semiparametric estimation. Section \ref{conclusion} concludes. Proofs and other technical material are collected in the appendix.

\section{Econometric Framework}
\label{framework}
This section presents the econometric framework for our paper. We are interested in estimating the binary choice model given by the data generating process
\begin{equation}\label{model1}
y^{*}_i  =  \mathbf{1}(\bm{x_i'\beta}_0 - \varepsilon^*_i>0 ) , 
\end{equation}
where $y^{*}_i  $ denotes true participation, $\bm{x_i}$ is a vector of observed characteristics with associated $k\times 1$ vector of coefficients of interest and $\varepsilon_i^*$ is the idiosyncratic error term which is assumed to be independently and identically distributed conditional on $\bm{x_i}$.\footnote{While we discuss the dependent variable in terms of program participation, note that our work applies to all binary choice models where the researcher is interested in a binary outcome.} Unfortunately, we do not observe true participation $y^{*}_i $ but rather the reported participation $y_i^{R}$ which is a possibly misclassified surrogate. \\ 

We can specify the relationship between $y^{R}_i $ and $y_i^{*} $ as

\begin{equation}\label{yR}
y^{R}_i  = y^*_{i}(1 - d_{i}) + (1 -  y^*_{i})d_{i}. 
\end{equation}

where $d_{i}\in\{0,1\}$ is an indicator variable representing the decision to misreport. In other words, misreporting occurs, i.e. $y^{R}_i= 1-y^*_{i}$ if an only if $d_i=1$, or equivalently, participation is correctly reported i.e. $y^{R}_i= y^*_{i}$ if and only if $d_i=0$.
This specification of a misreported indicator is quite general and has been considered by other studies such as \citeasnoun{denteh2022misclassification} to model misclassified treatments in difference-in-difference models. \citeasnoun{acerenza2021marginal} showed that this specification does not create any loss of generality.\footnote{Other authors such as \citet{jiang2020measurement, calvi2022women,tommasi2024bounding}, among others, have considered a specification of the form $y^{R}_i  = y^*_{i}d_{1i} + (1 -  y^*_{i})d_{0i}$, but \citeasnoun{acerenza2021marginal} showed that the binary nature of $y_i^R$ and $y_i^*$ imposes that $d_{1i}+d_{0i}=1$. Using a specification with different indicators for misclassification would not change our proposed method, but would make the exposition more cumbersome.} 
\\

The estimation of equation \eqref{model1}, given \eqref{yR} is challenging, especially when the misclassification probabilities are non-constant and potentially correlated with true participation. Studies that have attempted to provide estimators for equation \eqref{model1}, given \eqref{yR} have had to resort to simplifying assumptions such as constant misclassification probabilities \cite{hausman1998misclassification, abrevaya1999semiparametric,  keane2009classification} or one-sided misclassification \cite{nguimkeu2019estimation,lamarche2025quantile}. Other assumptions made in the literature are that the misclassification probabilities are uncorrelated with the probability of true participation \cite{bollinger1997modeling, meyer2017misclassification}. Some attempts have also been made to overcome these issues, chief among them being the possibility to obtain and use administrative data as a validation sample. However, obtaining such administrative data is difficult in practice. Moreover, recent evidence suggest that such administrative data may not be the so-called ``gold standard'' as commonly assumed \cite{courtemanche2019estimating}.\\ 

Fortunately, in some instances, the survey may have ``follow-up'' questions that allow the researcher to partially verify or validate the credibility of reported participation in a way that reduces the misclassification problem to some extent. The objective of this paper is to study such instances and propose solutions to permit credible estimation of binary choice models of program participation under  incomplete or one-sided verification. Suppose for instance that the survey has a follow-up question to the main program participation question that allows us to determine whether the individual truly participated in a program as they reported. In our setup, this means that we can verify reported participation for individuals with $y^R_i=1.$ We show that in these instances, the general (bi-directional) misclassification problem may be reduced to a uni-directional misclassification problem, therefore allowing for the identification of the true participation model using only the available survey data.\\

Formally, let $t_i$ denotes the truthfulness indicator of those that responded `Yes' for participation, with $t_i=1$ representing a correct report of participation and $t_i=0$ denoting a false report of participation. Note that,  unlike $d_i$ which is defined for all respondents,  $t_i$ is defined and observed only for those respondents who initially reported participation and is observed due to the follow-up verification question on that answer. Suppose we then complete the unobserved cases of $t_i$ by $0$'s and therefore define the variable $y_i$ by

\begin{equation}\label{PV}
y_i  = 
\begin{cases}
t_i & if ~ y^R=1  \\
 0 & if ~ y_i^R=0
\end{cases}
\end{equation}
That is, $y_i$ is the truthfulness of the status of those who reported participation and is set to $0$ for those who reported non-participation. Our first result immediately follows and states that under these conditions, the bi-directional misclassification problem reduces to a one-sided misclassification problem.

\begin{Lemma}[{\normalfont{Using partial verification to reduce the dimensionality of misclassification}}]\label{A1}\hfill\\
Suppose that self-reported participation, $ y^R_{i},$ is subject to misclassification and $y^*_{i}$  denotes the true, unobserved participation status. Let $d_i$ denotes the indicator of misreporting, and $t_i$ denotes the indicator of truthfulness of those who report program participation (i.e., those for whom $ y^R_{i}=1$).
Then, the partially validated participation variable, $y_i$, given by equation \eqref{PV} is characterized by  
\begin{equation}\label{ystar}
y_i = y_i^*(1-d_i)
\end{equation}

 \end{Lemma}
\begin{proof}
See Appendix.
\end{proof}

Equation \eqref{ystar} implies a misclassification model that preserves the true response, $y_i^*$, but differs from equation \eqref{yR} in important ways. While equation \eqref{yR} characterizes a two-way misclassification model, equation \eqref{ystar} is a one-way misclassification model. Essentially, if we use $y_i$ as a surrogate for true participation rather than the reported $y_i^R$, it only possibly suffers from errors of omission or false negatives (i.e., those whom $y_i=1$ are necessarily true participants, and only those whom $y_i=0$ could be misreporting). This means that estimating true program participation based on $y_i$ only requires to recover the false negatives rate, whereas estimating it based on $y^R_i$ may require estimating both the false negatives and the false positives rates. The latter could be substantially involving, especially when misclassification probabilities are correlated with true participation, and may not be identified unless under strong regularity conditions \citep{lewbel2000semiparametric}.\footnote{\citet{lewbel2000semiparametric} also explains that even with strong identification conditions, estimation in this general case is not likely to be very practical, because they involve up to third-order derivatives of the link function and repeated applications of nonparametric regression.}\\

Equation \eqref{ystar} corresponds to the partial observability model of \citet{poirier80}, which was utilized in a first stage estimation of a two-step treatment effects estimator proposed in \citet{nguimkeu2019estimation}, for situations where participation is misreported endogenously in one direction. In the next two sections, we first characterize the biases that arise from using either $y^R_i$ or $y_i$ as the outcome variable in equation \eqref{model1}, and propose an MLE estimator that consistently estimate the desired parameters $\bm{\beta}$, the participation probabilities $\esp[y_i^*|\bm{x}_i]$, as well as the marginal effects  $\partial\esp[y_i^*|\bm{x}_i]/\partial \bm{x}_i$ and their averages.

\section{Bias due to Misclassification} \label{biases}
\citet{meyer2017misclassification} discusses the resulting bias from using the possibly misclassified self-reported participation, $y^R_i$. Here, we additionally consider the bias that results from using the partially validated participation variable, $y_i$ as defined in equation \eqref{ystar} as a surrogate for the true participation, $y^*_i.$ We also formally characterize the asymptotic bias discussed in \citet{meyer2017misclassification} when self-reported participation $y_i^R$ is used as a surrogate for $y^*_i.$ This allows us to directly compare these biases and better understand the features of the associated estimators. Let $F(\cdot)$ denotes the cumulative distribution function (cdf) of the random variable $\varepsilon_i^*$ that we assumed to be continuously differentiable with probability density function (pdf) $f(\cdot)$. Recall that $\bm{\beta}_0$ is the true parameter vector of interest which is uniquely defined by
\[
\esp[y_i^*|\bm{x}_i]=\Pr[y_i^*=1|\bm{x}_i]=F(\bm{x}_i'\bm{\beta}_0).\\
\]

Suppose that the researcher knows $F(\cdot)$ and decides to estimate the binary choice model for program participation using the reported  participation $y_i^R$ as the outcome of interest, rather than the unobserved true participation $y_i^{*}$. The estimated model is given by
\begin{equation}\label{incorrectR}
P(y_i^R = 1|\bm{x_i})  = F(\bm{x}_i'\bm{\beta}). 
\end{equation} 
and the log-likelihood function of the model based on the incorrect specification in equation \eqref{incorrectR} is given by 
\[
\mathcal{L}_n^R(\beta)=\frac{1}{n}\sum_{i=1}^n \left\{y_i^R\ln F(\bm{x}_i'\bm{\beta})+(1-y_i^R)\ln \left(1- F(\bm{x}_i'\bm{\beta})\right)\right\}. 
\]

We define the covariate-dependent misclassification probabilities by 
\[
a_0^R(\bm{z}_i)=\Pr\left(y_i^R=1|y_i^{*}=0,\bm{z}_i\right)\equiv \text{probability of false positives}
\]
\[
a_1^R(\bm{z}_i)= \Pr\left(y_i^R=0|y_i^{*}=1,\bm{z}_i\right)\equiv \text{probability of false negatives}
\]
where $\bm{z}_i$ be an $l$-vector of covariates, $l\geq k$, that includes $\bm{x}_i$ so that misclassification probabilities may also depend on variables that possibly do not affect the true response. Denoting by $\widehat{\bm{\beta}}_R$ the maximum likelihood estimator of equation \eqref{incorrectR} and by $\bm{\beta}_R$ its probability limit, we have the following result.\footnote{Note that equation \eqref{biasR} does not imply a closed-form solution for $\bm{\beta}_R$ since both sides of the equation still have it. But it permits to write the bias in terms of components that allow to study its sign.}


\begin{theorem}[Bias using reported participation]\label{theoremR}\hfill\\
The maximum likelihood estimator, $\widehat{\bm{\beta}}_R$, of model \eqref{model1} using the reported participation, $y_i^R$, defined by \eqref{yR} rather than true participation, $y_i^{*}$, is inconsistent, and the asymptotic bias can be characterized as:
\begin{equation}\label{biasR}
\pl ~\widehat{\bm{\beta}}_R -{\bm{\beta}}_0=- \bm{A}_R^{-1}\esp\left[\left\{\frac{a_1^R(\bm{z}_i)}{1 - F(\bm{x}_i'\bm{\beta}_0)}-\frac{a_0^R(\bm{z}_i)}{ F(\bm{x}_i'\bm{\beta}_0)}\right\}f(\bm{x}_i'\bm{\beta}_0) \bm{x_i} \right]
\end{equation}
where $\bm{A}_R$ is a positive definite matrix defined by

\[
\bm{A}_R=\mathbb{E}\left[\left\{\bm{b}_i^R{\tilde{f}_i^{(1)}}+\bm{c}_i^R{\tilde{f}_i}^2\right\}\bm{x_{i}x_{i}}^\prime\right],
\]
\[
\text{with components:} \qquad \bm{b}_i^R=\frac{a_1^R(\bm{z}_i)}{1 - \tilde{F}_i}-\frac{a_0^R(\bm{z}_i)}{ \tilde{F}_i},
\]

\[
 \bm{c}_i^R=\frac{\tilde{F}_i(1-\tilde{F}_i)-(1-2\tilde{F}_i)\left[a_1^R(\bm{z}_i)\tilde{F}_i-a_0^R(\bm{z}_i)(1-\tilde{F}_i)\right]}{\tilde{F}_i^2(1-\tilde{F}_i)^2},
\]

$\tilde{F}_i=F(\bm{x_i'\tilde{\beta}})$,  and $\tilde{f}_i=f(\bm{x_i'\tilde{\beta}})$. Here,  $\tilde{f}^{(1)}$ is the derivative of $f(\cdot)$ evaluated at $\bm{x_i'\tilde{\beta}}$, 
and $\tilde{\bm{\beta}}$ lies on the line segment joining $\bm{\beta_{0}}$ and $\bm{\beta}_R$.
\end{theorem}
\begin{proof}
See Appendix.
\end{proof}
 
The asymptotic bias characterized here stems from the misclassification of the binary response and hence the misspecification of the associated binary choice model, as discussed by \citet{meyer2017misclassification}.  As explained by these authors, the sign of this bias cannot be predetermined. The above result shows that it would depend on the relative signs of the components in the vector of regressors, $\bm{x}_i$, on the one hand, as well as the difference between the ratio of misclassification probabilities (i.e. the risk ratio of misreporting)  and the ratio of true participants to true non-participants (i.e. the odds ratio of participation), $\dfrac{a_0^R(\bm{z}_i)}{a_1^R(\bm{z}_i)} -\dfrac{F(\bm{x}_i'\bm{\beta}_0)}{1 - F(\bm{x}_i'\bm{\beta}_0)}$. Even if we knew the sign of all components of the regression vector, the sign of the difference between the two ratios would still be unknown. In some exceptional cases the estimator will be consistent even if there is misclassification and misspecification. This would occur only if the risk ratio of misreporting and the odds ratio of participation are identical, but the researcher would not be able to verify this occurrence.\\

Now, suppose that the researcher decides to estimate the binary choice model of program participation using the partially validated participation, $y_i$, as the outcome of interest, in lieu of the unobserved true participation, $y_i^{*}$. The model they estimate is given by
\begin{equation}\label{incorrectP}
P(y_i = 1|\bm{x_i})  = F(\bm{x}_i'\bm{\beta}). 
\end{equation} 
and the log-likelihood function of the model based on the incorrect specification in equation \eqref{incorrectP} is given by 
\[
\mathcal{L}_n^P(\beta)=\frac{1}{n}\sum_{i=1}^n \left\{y_i\ln F(\bm{x}_i'\bm{\beta})+(1-y_i)\ln \left(1- F(\bm{x}_i'\bm{\beta})\right)\right\}. 
\]
 This model only has false negatives and their probability is 
\[
a_1(\bm{z}_i)= \Pr\left(y_i=0|y_i^{*}=1,\bm{z}_i\right).
\]
Denoting by $\widehat{\bm{\beta}}_P$ the maximum likelihood estimator of Model \eqref{incorrectP} and by $\bm{\beta}_P$ its probability limit, 
we have the following result.

\begin{theorem}[Bias using partially validated participation]\label{theoremP}\hfill\\
The maximum likelihood estimator, $\widehat{\bm{\beta}}_P$, of model \eqref{model1} using the partially validated participation, $y_i$, defined by \eqref{ystar} rather than true participation, $y_i^{*}$, is inconsistent and the asymptotic bias is characterized by:
\begin{equation}\label{biasP}
\pl ~\widehat{\bm{\beta}}_P -{\bm{\beta}}_0=-\bm{A}^{-1}\esp\left[\frac{a_1(\bm{z}_i)}{1 - F(\bm{x}_i'\bm{\beta}_0)}f(\bm{x}_i'\bm{\beta}_0) \bm{x_i} \right]
\end{equation}
where $\bm{A}$ is the positive definite matrix defined by

\[
\bm{A}=\mathbb{E}\left[\left\{\frac{a_1(\bm{z}_i)}{1 - \bar{F}_i}{\bar{f}_i^{(1)}}+\frac{(1-\bar{F}_i)-(1-2\bar{F}_i) a_1(\bm{z}_i)}{\bar{F}_i(1-\bar{F}_i)^2}{\bar{f}_i}^2\right\}\bm{x_{i}x_{i}}^\prime\right],
\]

$\bar{F}_i=F(\bm{x_i'\bar{\beta}})$,  $\bar{f}_i=f(\bm{x_i'\bar{\beta}})$, $\bar{f}^{(1)}$ is the derivative of $f(\cdot)$ evaluated at $\bm{x_i'\bar{\beta}}$, 
and $\bar{\bm{\beta}}$ lies on the line segment joining $\bm{\beta_{0}}$ and $\bm{\beta}_P$.\footnote{If normality is assumed for $\varepsilon_i$, then ${f}^{(1)}(\bm{x_i'\bar{\beta}})=-\bm{x_i'\bar{\beta}}\phi(\bm{x_i'\bar{\beta}})$, where $\phi(\cdot)$ is the pdf of the standard normal.}

\end{theorem}
\begin{proof}
See Appendix.
\end{proof}

The sign of the bias is generally unknown. However, since the matrix $\bm{A}$ is positive definite, the sign of the bias in a particular component of $\hat{\bm{\beta_P}},$ say $\hat{\bm{\beta}_P}^{(k)}$ will always be opposite to the sign of the corresponding random component, $x_{ik}$, if the latter has a constant sign.  In particular, the bias of any dummy explanatory variable will always be  negative (i.e., the associated coefficient will be underestimated). It is important to understand the substantive difference between the asymptotic bias derived in the case of one-sided misreporting here and the one implied by two-sided misreporting, as discussed in \citet{meyer2017misclassification}. As stated above, there are instances where naive estimator would be consistent, even in the presence of severe misreporting. This would arise, for example, if the normalized conditional distribution of the error term in the correct model happens to be the same as the misspecified  error distribution. This complicates the identification and estimation of the (two-sided misclassification) model in practice, especially in the situation where misclassification errors are person dependent and conditionally correlated with true participation. In contrast, in our one-sided misreporting model the naive estimator is always inconsistent, but in a way that can be identified as we discuss below.\\

We now provide MLE  approaches to consistently estimate $\bm{\beta}_0$ when  true participation $y_i^{*}$ is misclassified in both directions but we have access to a partially validated version, $y_i$, of  the self-reported participation variable $y_i^R$  as discussed earlier. We will distinguish the cases where the cumulative distribution function of the true participation, $F(\cdot)$ is known and when it is unknown. 

\section{Proposed Maximum Likelihood Estimators} \label{proposed-MLE}
Recall that from equation (\ref{yR}), we have $y_i^R=y_i^*(1-d_i)+(1-y_i^*)d_i$, which implies $y_i^*y_i^R=y_i^*y_i^*(1-d_i)+y_i^*(1-y_i^*)d_i=y_i^*(1-d_i)$. Hence, by equation \eqref{ystar}, the correctly specified model defined with partially validated participation can also be written as

\begin{equation}\label{newmodel}
y_i = y_i^* y_i^R
\end{equation}

where
$y^{*}_i  =  \mathbf{1}(\bm{x_i'\beta} - \varepsilon^{*}_i >0)$ as specified earlier in Equation \eqref{model1}. We specify the decision to misreport as 
\begin{equation}\label{model2}
d_i  =  \mathbf{1}\left(\bm{z_i'\beta}_R - \varepsilon_i<0\right)
\end{equation}
where $\bm{z_i}$ is a vector of covariates that possibly contains and/or overlaps with $\bm{x_i}$. Our estimation of the main quantities of interests stated earlier, that is, $\bm{\beta}$, $\esp[y_i^*|\bm{x_i}]$ and $\partial\esp[y_i^*|\bm{x_i}]\partial \bm{x_i}$ is therefore based on the model defined by equation \eqref{newmodel}, where $y^{*}_i$ and $d_i$ are specified by equations (\ref{ystar}) and (\ref{model2}), respectively.\\ 

Equation \eqref{newmodel} resembles \citet{poirier80}'s partial observability model. However, in our case, we are able to observe more than in Poirer's model because $y_i^R$ is observed, but less than in a full observability case, because $y_i^*$ is only sometimes observed. In particular, we observe $y_i^*$ if and only if $y_i^R=1$. But if $y_i^R=0$, then we have no information about $y_i^*$, implying that $y_i^*$ could assume the value 1 (for errors of omission or false negatives) or 0 (for true non-participants). In other words, the two pairs  ($y_i^R=0 ~, ~y_i^*=1$) and ($y_i^R=0 ~, ~ y_i^*=0$) are the two indistinguishable scenarios when $y^R_i=0$. This is therefore close to a type of censored Probit which \citet{meng1985cost} referred to as ``Partial Partial Observability'' to distinguish it from from ``Partial Observability'' in the sense of \citet{poirier80}.\\  

However, the specification in equation \eqref{newmodel} is still different from \citet{meng1985cost} and from the traditional censored Probit model due to the complexity of $y_i^R$, given that it is misclassified as specified by equation \eqref{yR}. \textcolor{black}{Based on the above models, we propose two maximum likelihood estimators. The first estimator is based on the partial observability model (PO MLE) given by Equation \eqref{ystar} as discussed in the first-step of \citeasnoun{nguimkeu2019estimation}'s procedure. The second estimator uses equation \eqref{newmodel} and can be thought of as a Partial Partial Observability MLE (PPO MLE) given that $y_i^R$ is observed. The advantage of the latter is the {potential efficiency} gain from the extra piece of information obtained by observing one of the selection indicators, $y_i^R$, on the right hand side of equation \eqref{newmodel}, whereas for equation \eqref{ystar} none of the right hand side indicators is observed. Although the MLE techniques employed are not new themselves, the novelty in the proposed methods lie in the way they can be used to solve the two-sided misclassification problem in a binary choice model when the accuracy of only one category in the outcome can be partially validated.}\\

These parametric maximum likelihood methods are considered for the case of an arbitrary bivariate distribution of $\varepsilon_i^*$ and $\varepsilon_i$, and possibly different vector of regressors $\bm{x_i}$ and $\bm{z_i}$, with the following assumptions.

\begin{Assumption}[{Independence}]\label{ml1}\hfill\\ 
The error terms $(\varepsilon_i^*, \varepsilon_i)$ are independent of the exogenous variables $(\bm{x_i},\bm{z_i})$ and are identically distributed with a joint CDF denoted $G(\varepsilon,\varepsilon^*)=\Pr[\varepsilon_i<\varepsilon,\ \varepsilon_i^*<\varepsilon^*]$. 
\end{Assumption}

The CDF, $G(. , .),$ is assumed to be sufficiently flexible to capture plausible forms of dependence between $\varepsilon$ and $\varepsilon^*$. However, we impose two restrictions on $G(\cdot,\cdot)$. First, we require $G(\cdot,\cdot)$ to have identical marginal distributions for $\varepsilon$ and $\varepsilon^*$. This doesn't create any substantial loss of generality given that the location and scale parameters for $\varepsilon$ and $\varepsilon^*$ can usually be normalized by applying the appropriate shifts in $y_i^R$ and $y_i^*$ as in the usual binary choice situations. However, it is useful to derive analytical results that are less cumbersome. Second, we will restrict ourselves to one-parameter families of  $G(\cdot,\cdot)$ denoted by $\rho$. In the case of Normality, this parameter would be the correlation coefficient between $\varepsilon_i$ and $\varepsilon_i^*$. In other types of distributions, for example Logit, the interpretation of such a parameter is a bit more sophisticated than just correlation. Formally, the first  and second restrictions on $G(\cdot,\cdot)$ can be summarized as:
\begin{enumerate}[(i)]
	\item $G(\varepsilon,\varepsilon^*)=G(\varepsilon,\varepsilon^*;\rho)$ depends on $\rho$
	 \item $F(\varepsilon)=G(\varepsilon, \infty;\rho)$, $F(\varepsilon^*)=G(\infty, \varepsilon^*;\rho)$, and $F(\cdot)$ does not depend on $\rho$.
\end{enumerate}

\begin{Assumption}[{Identification}]\label{ml2}\hfill\\ 
The variables $\bm{x_i}$ and $\bm{z_i}$ are full rank, both have at least one continuous covariate, 
and at least one covariate in $\bm{z_i}$ is not relevant for $\bm{x_i}$ (or vice versa). 
\end{Assumption}

This means that there should be no perfect collinearity in the exogenous regressors. The additional condition implies that there should also be at least one exclusion-restriction between these regressors. As explained by previous authors \citeaffixed{poirier80, meng1985cost, greene2003econometric}{e.g.}, these conditions are sufficient to guarantee that the associated parameters are (locally) identified since they imply that the information matrices in the likelihood maximization problems that are considered will be nonsingular. Following \citeasnoun{nguimkeu2019estimation}, we are thus relying on the general principle of \citeasnoun{Rothenberg1971} for the parameter identification of our model. We recognize that the exclusion-restriction between the covariates $\bm{x_i}$ of true participation and the covariates  $\bm{z_i}$ of the mismeasured reported participation may be difficult to obtain in practice. Our suggestion for exclusion-restrictions in $\bm{z_i}$ is to use factors that could specifically affect the reporting process without affecting the true participation such as peculiar features of the survey and its administration (e.g., date, length, time or mode of the interview).\\    

\textcolor{black}{From this framework, the covariate-dependent probability of false positives and false negatives can be expressed as 
\[
a_0^R(\bm{z}_i,\bm{x}_i)=\Pr\left(y_i^R=1|y_i^{*}=0,\bm{z}_i,\bm{x}_i\right)=\dfrac{G\left(\bm{z_i^\prime\beta}_R,\, \bm{x_i^\prime\beta}, \rho\right)}{F(\bm{x_i'{\beta}})},
\]
and
\[
a_1^R(\bm{z}_i,\bm{x}_i)= \Pr\left(y_i^R=0|y_i^{*}=1,\bm{z}_i,\bm{x}_i\right)=\dfrac{F(\bm{x_i'{\beta}})-G\left(\bm{z_i^\prime\beta}_R,\, \bm{x_i^\prime\beta}, \rho\right)}{1-F(\bm{x_i'{\beta}})},
\]
respectively. Their consistent estimators are obtained by plugging a vector of consistent parameter estimates into the formulas. These general misclassification probability functions account for both covariate-dependent and possibly endogenous misclassification errors.\\  
}

If we denote by $\bm{\hat{\beta}}$ a consistent estimators of  $\bm{{\beta}}$  obtained, e.g., from the method discussed below, the predicted true participation is then estimated by
\[
\widehat{\esp}[y_i^*|\bm{x_i}]=\widehat{y_i}^* = F(\bm{x_i'\hat{\beta}})
\]
and the estimated marginal effect of the $j^{th}$ covariate of $\bm{x_i}$ on true participation is given by
\[
\dfrac{\partial \widehat{\esp}[y_i^*|\bm{x_i}]}{\partial x_{ij}} = f(\bm{x_i'\hat{\beta}})\hat{\beta}_j
\]
where $F(\cdot)$ and $f(\cdot)$ are the assumed (or estimated) CDF and PDF of $\varepsilon_i^*$, as we discuss later, and  $\hat{\beta}_j$ is the $j^{th}$ component of $\bm{\hat{\beta}}$. The average marginal effect is obtained by taking the average of the above quantity over the whole sample.\\


Estimators similar to the proposed PPO MLE have appeared in works such as \citet{farber1981worker,connolly1984impact,meng1985cost,boyes1989econometric,butler1996estimating}, and \citet{feinstein1990detection}. However, all of these existing ones were considered under a much simpler form of $y_i^R$, under the normality assumption, and sometimes under the assumption of a zero conditional correlation between the random components in the specification of the indicators on the right hand side of Equation \eqref{newmodel}.
We derive the PPO MLE here under a misclassified $y_i^R$ which means a much more complex form, and under more general conditions regarding the regressors and the error distributions, all of which may have important consequences for identification.

\subsection{Likelihood Function of the PPO MLE Estimator}
\textcolor{black}{For the PPO MLE,} we calculate the probabilities of the three possibly distinguishable cases:  a successfully validated participation $(y_i^R=1,\ y_i^*=1)$,  an unsuccessfully validated participation $(y_i^R=1,\ y_i^*=0)$,  and an unverified non-participation $(y_i^R=0)$.
Denote by $H(\cdot,\cdot,\rho)$ the upper tail probability of $G(\cdot,\cdot,\rho)$. That is, 
\begin{equation*}
H(\varepsilon,\varepsilon^*,\rho)= \Pr[\varepsilon_i>\varepsilon,\ \varepsilon_i^*>\varepsilon^*]=1-F(\varepsilon)-F(\varepsilon^*)+G(\varepsilon,\varepsilon^*,\rho)
\end{equation*}
Then the probability of a successfully validated participation is: 
\begin{align*}
\small
P_i(\bm{\beta}_R,\, \bm{\beta,} \rho) &= \Pr[y_i^R=1,y_i^*=1|\bm{x_i, z_i}]=\Pr[d_i=0,y_i^*=1|\bm{x_i, z_i}]\\\
   &=\Pr[\varepsilon_i<\bm{z_i^\prime\beta}_R,\,  \varepsilon_i^*<\bm{x_i^\prime\beta}]=G\left(\bm{z_i^\prime\beta}_R,\, \bm{x_i^\prime\beta}, \rho\right)\numberthis \label{success}
\end{align*}
Also, the probability of unsuccessfully validated participation is
\begin{align}\label{failed}
\small 
Q_i(\bm{\beta}_R,\, \bm{\beta,} \rho) &= \Pr[y_i^R=1,y_i^*=0|\bm{x_i, z_i}]=\Pr[d_i=1,y_i^*=0|\bm{x_i, z_i}]\\
   &=\Pr[\varepsilon_i>\bm{z_i^\prime\beta}_R,\,  \varepsilon_i^*>\bm{x_i^\prime\beta}]=H\left(\bm{z_i^\prime\beta}_R,\,  \bm{x_i^\prime\beta}, \rho\right)\numberthis \label{success}
\end{align}

Finally, the probability of unverified non-participation is:
\begin{equation}\label{unverified}
\small
\Pr[y_i^R=0|\bm{x_i, z_i}]=1-P_i(\bm{\beta}_R,\bm{\beta,} \rho)-Q_i(\bm{\beta}_R,\bm{\beta,} \rho)
\end{equation}

Combining these formulas, we obtain the likelihood function as:
\begin{align*}
\small
\mathcal{L}_n(\bm{\beta}_R,\bm{\beta}, \rho)&=\sum_{i=1}^{n}\bigg[y_i^R y_i^* \ln{P_i(\bm{\beta}_R,\bm{\beta}, \rho)} + y_i^R(1-y_i^*)\ln{Q_i(\bm{\beta}_R,\bm{\beta}, \rho)}\\
& + (1-y_i^R)\ln{[1-P_i(\bm{\beta}_R,\bm{\beta,} \rho)-Q_i(\bm{\beta}_R,\bm{\beta,} \rho)]\bigg]}
\end{align*}
Recalling that $y_i^Ry_i^*=y_i$ and noticing that $y_i^R(1-y_i^*)=y_i^R-y_i$, the likelihood function can be rewritten as:
\begin{equation}\label{mlelike}
\begin{aligned}
\small
\mathcal{L}_n(\bm{\beta}_R,\bm{\beta}, \rho)&=\sum_{i=1}^{n}\bigg[y_i \ln{P_i(\bm{\beta}_R,\bm{\beta}, \rho)} + (y_i^R-y_i)\ln{Q_i(\bm{\beta}_R,\bm{\beta}, \rho)}\\
& + (1-y_i^R)\ln{[1-P_i(\bm{\beta}_R,\bm{\beta,} \rho)-Q_i(\bm{\beta}_R,\bm{\beta,} \rho)]\bigg]}=\sum_{i=1}^n\mathcal{L}_i(\bm{\theta})
\end{aligned}
\end{equation}

The MLE of the parameter vector $\bm{\theta}=(\bm{\beta}_R,\bm{\beta,} \rho)$ is obtained by maximizing the log-likelihood function \eqref{mlelike} with respect to the components of $\bm{\theta}$, which can be derived by numerically solving the first order conditions given by $\sum_{i=1}^nS_i(\bm{\theta})=\bm{0}$, where 
$S_i(\bm{\theta})$ is the individual score vector defined by
$S_i(\bm{\theta})=S_i(\bm{\beta}_R,\bm{\beta}, \rho) =\dfrac{\partial \mathcal{L}_i(\bm{\theta})}{\partial\bm{\theta}}.$\\

Under the regularity conditions stated above, the derived MLE, $\bm{\hat{\theta}}=(\bm{\hat\beta}_R,\bm{\hat\beta},\hat\rho)$, is consistent and asymptotically normal, such that $\sqrt{n}(\bm{\hat{\theta}-\theta})\cvd N\left(\bm{0},\ \bm{\mathcal{I}(\bm{\theta})}^{-1}\right)$,
where
$\bm{\mathcal{I}(\bm{\theta})}=\esp\left[S_i(\bm{\theta})S_i(\bm{\theta})'\right]$ is the information matrix. A consistent estimator of the asymptotic variance of the MLE $\bm{\hat{\theta}}=(\bm{\hat\beta}_R,\bm{\hat\beta},\hat\rho)$ can then be obtained as $\bm{\mathcal{I}(\bm{\hat\theta})}^{-1}/n$, where
\[
\bm{\mathcal{I}(\bm{\hat\theta})}=\mathbb{E}_n\left[S_i(\bm{\hat\theta})S_i(\bm{\hat\theta})'\right],
\]

and $\mathbb{E}_n[\cdot]$ denotes the sample average operator. This result can be easily applied to the common particular cases where error terms have the normal or logistic distributions. \\

First, for the normal case, we have $G(\varepsilon,\varepsilon^*;\rho)=\Phi_2(\varepsilon,\varepsilon^*;\rho)$ such that:
\[
P_i(\bm{\beta}_R,\bm{\beta},\rho)=\Phi_2(\bm{z_i^\prime\beta}_R,\,\bm{x_i^\prime\beta};\, \rho),\, \quad  \text{and}\, \quad  Q_i(\bm{\beta}_R,\bm{\beta},\rho)=\Phi_2(-\bm{z_i^\prime\beta}_R,\,-\bm{x_i^\prime\beta};\, \rho), 
\]

where  $\Phi_2(\cdot,\cdot;\rho)$ is the CDF of the bivariate standard normal with correlation $\rho$. If $\rho=0$, that is, there is no correlation between the reported and the true participation conditional on the covariates, then we have
\[
Q_i(\bm{\beta}_R,\bm{\beta},0)=\Phi_2(-\bm{z_i^\prime\beta}_R,\,-\bm{x_i^\prime\beta};\, 0)=\left(1- \Phi(\bm{z_i^\prime\beta}_R)\right)\left(1-\Phi(\bm{x_i^\prime\beta})\right)
\] 
where  $\Phi(\cdot)$ is the CDF of the standard normal. The partial derivatives for the above bivariate integrals are given in Section \ref{simulations} below  \citeaffixed{drezner1990computation,greene2003econometric}{see also,}. \\

Second, for the logistic case, there are various options for choosing the bivariate distribution. Many suggestions have been proposed in the literature with restrictive forms of correlations, such as \citeasnoun{Gumbel1961}, \citeasnoun{johnson1972continuous}, \citeasnoun{ray1980correcting}, \citeasnoun{dubin1989selection}. We propose to use the bivariate distribution of \citeasnoun{ali1978class} which is indexed to an unknown parameter $\rho$ that is subject to the restriction $-1<\rho<1$, similar to a correlation coefficient. This bivariate logistic distribution is defined by

\[
G(\varepsilon,\varepsilon^*;\rho)=\dfrac{\Lambda(\varepsilon)\Lambda(\varepsilon^*)}{\big[1-\rho (1-\Lambda(\varepsilon))(1-\Lambda(\varepsilon^*))\big]}=\dfrac{1}{1+e^{-\varepsilon}+e^{-\varepsilon^*}+(1-\rho)e^{-\varepsilon-\varepsilon^*}}
\]
where $\Lambda(t)=[1+e^{-t}]^{-1}$ is the CDF of the logistic distribution. 

Note that while $\rho$ is not exactly the correlation coefficient, it is a measure of association between $\varepsilon$ and $\varepsilon^*$, and bears similar properties as the usual correlation. In particular, if $\rho=0$ then $G(\varepsilon,\varepsilon^*;0)=\Lambda(\varepsilon)\Lambda(\varepsilon^*)$ and the two error terms are independent.
We therefore have

\[
P_i(\bm{\beta}_R,\bm{\beta},\rho)=\dfrac{\Lambda(\bm{z_i^\prime\beta}_R)\Lambda(\bm{x_i^\prime\beta})}{[1-\rho (1-\Lambda(\bm{z_i^\prime\beta}_R))(1-\Lambda(\bm{x_i^\prime\beta}))]} \qquad \text{and}\, 
\]

\[
\quad Q_i(\bm{\beta}_R,\bm{\beta},\rho)=1-\Lambda(\bm{z_i^\prime\beta}_R)-\Lambda(\bm{x_i^\prime\beta})+\dfrac{\Lambda(\bm{z_i^\prime\beta}_R)\Lambda(\bm{x_i^\prime\beta})}{\left[1-\rho (1-\Lambda(\bm{z_i^\prime\beta}_R))(1-\Lambda(\bm{x_i^\prime\beta}))\right]}
\]

As with the normal case, when $\rho=0$ there is no correlation between the reported and the true participation conditional on the covariates, so that the joint probabilities and just the product of marginal probabilities. These probabilities are somewhat easier to compute than the normal case given that no integration is involved, and their derivatives are relatively straightforward but tedious.

\subsection{Likelihood Function of the PO MLE Estimator}
For the PO MLE, we know the respondent truly participated if and only if their participation was successfully verified $(y_i=1~ \text{if and only if} ~y_i^* =y_i^R=1)$ and we observe $y_i=0$ otherwise. This observation mechanism follows from Lemma \ref{A1} where $y_i=y_i^*(1-d_i)$ and $d_i$ is the latent indicator of misreporting. Here, for $y_i=0$, we we cannot distinguish between those who did not truly participate ($y_i^*=0$) and false negatives (those who truly participated but are misclassified in the data). The log-likelihood function for this estimator is therefore
\[
\mathcal{L}_n(\bm{\beta}_R,\bm{\beta}, \rho)=\sum_{i=1}^{n}\bigg[y_i \ln{P_i(\bm{\beta}_R,\bm{\beta}, \rho)} +(1-y_i) \ln{\left[1-P_i(\bm{\beta}_R,\bm{\beta}, \rho)\right]}\bigg]
\]
Details about the identification and optimization, including the consistency and asymptotic normality of the corresponding estimator can be found in \citet{nguimkeu2019estimation} (see first step of their two-step procedure discussed on page 492). As explained earlier, common particular cases such as the normal case and the logistic case can be considered by appropriately defining the joint CDF $P_i(\bm{\beta}_R,\bm{\beta}, \rho)$ as given above. The PO MLE obtained here will generally be less efficient than the PPO MLE derived above, given the relatively lower level of observability (i.e., less information), especially when the data are noisier (i.e., high misclassification rates). Unfortunately, as noted by \citet{poirier80}(see page 212), quantifying the efficiency loss is not possible without referring to a particular data set \citeaffixed{meng1985cost}{see also}. This means that we can only compare their efficiencies in a Monte Carlo simulation or other types of data examples. In the following section, we perform Monte Carlo simulations.

\section{Monte Carlo simulations}\label{simulations}

This section presents the results of Monte Carlo simulations, comparing the proposed estimators with standard approaches that ignore misclassification or assume constant misclassification probabilities. We evaluate performance under non-random covariate-dependent misclassification scenarios, examining how partial verification affects estimation of binary choice models across different average misreporting rates.

\subsection{Simulation Setup}

The baseline data generating process is simulated as follows. The true outcome (participation) indicator, $y_i^*$, is given by
\begin{equation}
y_i^* = \mathbf{1}(\mathbf{x}_i'\boldsymbol{\beta}_0 - \varepsilon_i^* > 0)
\end{equation}
where $\mathbf{x}_i$ is a $4 \times 1$ vector of covariates with associated coefficient vector $\boldsymbol{\beta}_0 = [2, -0.5, 0.5, 1]'$. These covariates are drawn from various normally distributed random variables. The decision to misreport follows
\begin{equation}
d_i = \mathbf{1}(\mathbf{z}_i'\boldsymbol{\beta}_R - \varepsilon_i < \tau), 
\end{equation}
where $\mathbf{z}_i$ contains the first three covariates in $\mathbf{x}_i$ and a fourth covariate that is excluded from $\mathbf{x}_i$ (also normally distributed), and $\boldsymbol{\beta}_R = [1, 0.5, -1.5, 2.5, 1]'$. The threshold $\tau$ is calibrated to achieve desired \emph{average} misreporting rates. The error terms $(\varepsilon_i^*, \varepsilon_i)$ follow a bivariate normal distribution with unit variances and correlation $\rho = -0.8$. We allow misclassification probabilities to be dependent on covariates (i.e., conditionally non-random) and unobserved factors potentially correlated with true participation through the error correlation structure. The reported participation is generated as $y_i^R = y_i^*(1-d_i) + (1-y_i^*)d_i$, representing two-sided misclassification. Based on Lemma \ref{A1}, partial validation eliminates false positives, yielding the partial validated outcome (participation) variable $y_i = y_i^*(1-d_i)=y_i^*y_i^R$, which exhibits only false negatives.

\begin{table}[htp!]
	\footnotesize
	\centering
	\caption{\textbf{Simulation Results}}
	\label{tab1}
	\begin{threeparttable}		
		
\begin{tabular}{@{}P{1.5cm}P{1.5cm}P{1cm}P{1.8cm}P{1.8cm}P{2cm}P{1.5cm}P{1.5cm}P{1.7cm}@{}}
\toprule
Average false negative rate & Average false positive rate & True $\beta_0$ & \multicolumn{4}{P{7.4cm}}{Naive estimators} &  \multicolumn{2}{P{3cm}}{Proposed Estimators} \\ 
\cmidrule{4-9} \\ 
 &  &  &   Probit \newline (Reported) & Probit \newline (Validated) & Probit \newline (Restricted) & HAS & PO MLE & PPO MLE \\ 
(1) &(2) &  (3) & (4)  & (5) &(6) & (7) & (8) & (9)  \\ 
 \hline 
\multirow{8}{*}{5\%}  & \multirow{4}{*}{5\%}  & 2    & 0.938  & 1.776  & 1.256  & 1.861  & 2.054  & 1.972  \\
                      &                       & -0.5 & -0.199 & -0.579 & -0.649 & -0.551 & -0.519 & -0.550 \\
                      &                       & 0.5  & 0.172  & 0.682  & 0.871  & 0.613  & 0.560  & 0.668  \\
                      &                       & 1    & 0.475  & 0.863  & 0.545  & 0.915  & 1.061  & 0.960  \\ \cmidrule{2-9}
                      & \multirow{4}{*}{20\%} & 2    & 0.599  & 1.782  & 1.427  & 1.856  & 2.057  & 2.022  \\ 
                      &                       & -0.5 & -0.024 & -0.579 & -0.704 & -0.551 & -0.518 & -0.488 \\
                      &                       & 0.5  & -0.068 & 0.677  & 0.984  & 0.614  & 0.556  & 0.671  \\
                      &                       & 1    & 0.322  & 0.865  & 0.660  & 0.910  & 1.062  & 1.016  \\ \cmidrule{1-9}
\multirow{8}{*}{10\%} & \multirow{4}{*}{5\%}  & 2    & 0.840  & 1.627  & 1.205  & 1.723  & 2.127  & 1.868  \\
                      &                       & -0.5 & -0.212 & -0.610 & -0.676 & -0.580 & -0.549 & -0.582 \\
                      &                       & 0.5  & 0.212  & 0.761  & 0.891  & 0.687  & 0.645  & 0.754  \\
                      &                       & 1    & 0.421  & 0.776  & 0.531  & 0.835  & 1.090  & 0.890  \\ \cmidrule{2-9}
                      & \multirow{4}{*}{20\%} & 2    & 0.520  & 1.622  & 1.418  & 1.724  & 2.139  & 1.939  \\
                      &                       & -0.5 & -0.036 & -0.611 & -0.734 & -0.579 & -0.554 & -0.472 \\
                      &                       & 0.5  & -0.032 & 0.758  & 1.038  & 0.691  & 0.644  & 0.757  \\
                      &                       & 1    & 0.277  & 0.775  & 0.643  & 0.834  & 1.092  & 0.954  \\ \cmidrule{1-9}
\multirow{8}{*}{20\%} & \multirow{4}{*}{5\%}  & 2    & 0.722  & 1.442  & 1.211  & 1.511  & 2.168  & 1.705  \\
                      &                       & -0.5 & -0.220 & -0.636 & -0.702 & -0.601 & -0.585 & -0.615 \\
                      &                       & 0.5  & 0.246  & 0.834  & 0.961  & 0.766  & 0.850  & 0.721  \\
                      &                       & 1    & 0.353  & 0.669  & 0.513  & 0.718  & 1.049  & 0.743  \\ \cmidrule{2-9}
                      & \multirow{4}{*}{20\%} & 2    & 0.425  & 1.434  & 1.368  & 1.524  & 2.169  & 1.794  \\
                      &                       & -0.5 & -0.042 & -0.634 & -0.772 & -0.607 & -0.580 & -0.468 \\
                      &                       & 0.5  & 0.000  & 0.843  & 1.081  & 0.773  & 0.857  & 0.737  \\
                      &                       & 1    & 0.224  & 0.670  & 0.618  & 0.717  & 1.057  & 0.817  \\
\bottomrule 
\end{tabular}

		\begin{tablenotes}[flushleft]
			\footnotesize
			\item \emph{Notes.} Columns 1 and 2 report average misclassification rates. Column 3 shows the true coefficients. Column 4 reports the MLE estimates using the originally measured outcome variable. Column 5 reports the MLE estimates using the partially validated outcome variable. Column 6 uses same method in Column 5 on a restricted of those whose responses were validated. Column 7 reports results from the \citet{hausman1998misclassification} approach. Column 8 reports the  results from the PO MLE  approach and Column 9 reports the results from the PPO MLE estimator described in Section \ref{proposed-MLE}.
		\end{tablenotes}
	\end{threeparttable}
\end{table}

\begin{table}[htp!]
	\footnotesize
	\centering
	\caption{\textbf{Comparison of Finite-Sample Efficiency: PO vs PPO}}
	\label{tab_efficiency}
	\begin{threeparttable}		
		\begin{tabular}{@{}P{2.2cm}P{2.2cm}ccccc@{}}
\toprule
\multicolumn{2}{c}{Error Rates} & \multicolumn{5}{c}{Ratio of Variances and Determinants (PO / PPO)} \\
\cmidrule(r){1-2} \cmidrule(l){3-7}
Average False Negative Rate & Average False Positive Rate & Var($\beta_1$) & Var($\beta_2$) & Var($\beta_3$) & Var($\beta_4$) & Det($\Sigma$) \\ 
\midrule
\multirow{2}{*}{5\%} & 5\% & 0.011 & 0.105 & 1.207 & 0.059 & 0.007 \\
 & 20\% & 0.053 & 2.434 & 0.737 & 0.190 & 0.047 \\ 
\cmidrule{1-7}
\multirow{2}{*}{10\%} & 5\% & 0.042 & 0.464 & 1.395 & 0.372 & 0.059 \\
 & 20\% & 1.378 & 1.617 & 1.304 & 1.498 & 5.287 \\ 
\cmidrule{1-7}
\multirow{2}{*}{20\%} & 5\% & 1.560 & 2.242 & 1.453 & 1.945 & 22.470 \\
 & 20\% & 1.805 & 2.387 & 2.070 & 2.230 & 29.000 \\ 
\bottomrule
\end{tabular}
		\begin{tablenotes}[flushleft]
			\footnotesize
			\item \emph{Notes.} Table compares the finite-sample efficiency of the the PO MLE and the PPO MLE estimators described. Columns 1 and 2 report average misclassification rates. 
            Each cell reports the Monte Carlo estimated variance ratio Var(PO)/Var(PPO) for the corresponding coefficient, and the determinant ratio Det(PO)/Det(PPO) for the estimated covariance matrix. Ratios exceeding one indicate that PPO yields higher efficiency than PO.
		\end{tablenotes}
	\end{threeparttable}
\end{table}

\subsection{Simulation Results}

We report simulation results averaged over 250 replications, each with sample size 5,000, for misclassification scenarios that yield different average false negative rates of approximately 5\%, 10\%, and 20\% as well as average false positive rates of approximately 5\% and 20\%. We compare our proposed estimators, the Partial observability Maximum Likelihood Estimator  (PO MLE) and the Partial Partial Observability Maximum Likelihood Estimator (PPO MLE) with several existing approaches. The naive estimators include the standard MLE based on the initial misclassified participation variable $y_i^R$, the standard MLE using the partially validated participation variable $y_i$, the standard MLE using the restricted sample of those subject to partial validation $t_i$ (i.e. those who self-reported in the affirmative to have participated in the program), and the \citet{hausman1998misclassification} method (denoted HAS in the tables of results). In our framework, note that only PO MLE and PPO MLE estimators are consistent for the true parameters in the presence of covariate-dependent misclassification.\\ 

Table \ref{tab1} shows that misclassification creates severe bias in existing methods. If the researcher completely ignores misclassification and estimate standard MLE using the initial measured participation variable, Column 4 shows that this approach leads to severe biases, including sign-switching in some instances. Column 5 shows that performing standard MLE using the partially validated participation variable continues to show severe attenuation bias, although the signs are preserved in our simulation study. Although it may be tempting to conduct MLE on the restricted sample of respondents that were subject to (partial) validation, simulations confirm that doing so results in substantial bias in our framework (Column 6). Also, we find that the \citet{hausman1998misclassification} approach also performs poorly, especially at higher levels of misclassification (Column 7). This finding is expected given that \citet{hausman1998misclassification} assumes that misclassification probabilities are random conditional on covariates. Finally, Columns 8 and 9 report results for the two consistent estimators proposed in our framework---the PO MLE and the PPO MLE estimators described in Section \ref{proposed-MLE}. We find that both approaches perform well across different misreporting rates, although there are differences in efficiency.\\

{Table \ref{tab_efficiency} reports simulation results examining the relative performance of the PO and PPO estimators by comparing their finite-sample efficiency across the same misclassification regimes. For each design, we estimate the variance of each estimator's components, $\Var(\beta_1), \Var(\beta_2), \Var(\beta_3),\Var(\beta_4)$, as well as the determinant of the estimated covariance matrix, $\det(\Sigma)$, to summarize the overall dispersion of each estimator. We report the ratios of these variances and determinants, where values above 1 indicate greater efficiency of the PPO estimator compared to the PO estimator. When misclassification rates are modest, the PO estimator is slightly more efficient than the PPO estimator in out setup (the ratios are all less than 1). However, as misclassification increases, the efficiency losses of PO become substantial. At average false negative rates of 10\% or 20\%, the variance ratios for most coefficients exceed one, and the determinant ratios grow dramatically. These results reinforce the intuition that while PO and PPO are consistent, PPO is more likely to fully leverage the additional information provided through partial validation.}\\ 

These simulation results provide evidence that partial validation, when combined with appropriate misclassification models, can address the estimation challenges posed by misclassified binary variables. In particular, the results suggest that approaches based on partial observability can overcome the challenges of estimating binary choice models with a misclassified outcome variable.

\section{Empirical Illustration: Health Insurance in Ghana} \label{empirics}

To illustrate the practical relevance of our methodology, we investigate the take-up of health insurance in Ghana, accounting for the potential misreporting of insurance coverage (i.e., the NHIS) using the Ghana Demographic and Health Survey. This empirical application provides an ideal setting to illustrate the importance of our framework for estimating binary choice models, as health insurance status is frequently misreported in survey data and the Ghana-DHS includes the partial validation structure that motivates our approach. We use data from the 2008 wave of the Ghana-DHS, which interviewed a total of 12,000 households. Following \citet{asare2020health}, our analysis focuses on women aged 15-49 who were asked detailed questions about their participation in Ghana's NHIS. The final analytic sample includes 3,187 women with non-missing information on all the included covariates.\\

The Ghana-DHS respondents are first asked about their NHIS participation status. Crucially, those who report having NHIS coverage are then asked to produce their NHIS identity cards for verification by the interviewer. This creates a partial validation structure where reported participation can be verified for those claiming coverage, but non-participants are not subject to any validation process. As established in our theoretical analysis, this transforms the original two-sided misclassification problem into a one-sided misclassification problem where only false negatives remain. Our outcome of interest is NHIS participation status, which we observe in two forms in the data---self-reported participation, and partially validated participation (those who can produce cards). Note that true participation status, which is of interest, is not observed in the data. Based on administrative data from the National Health Insurance Authority, almost 40\% of Ghanaians participated in NHIS in 2014 \citep{asare2020health}. Among respondents who initially claimed NHIS coverage, 40\% could not produce valid NHIS cards during the interview, suggesting substantial over-reporting of insurance coverage. Moreover, in our 2014 Ghana-DHS sample, 59\% of respondents reported NHIS coverage, consistent with an over-reporting of coverage. This finding is aligned with social desirability biases leading to misreporting of participation in government programs.\\

Table \ref{empirical_results} presents our main estimation results (coefficient estimates) examining the determinants of (true) NHIS participation given various covariates, including age, household wealth, rural residence, number of children, education, religion, and ethnicity. All estimations cluster standard errors at the level of district of residence. In addition, we include an exclusion restriction in the true participation, which is the number of years of exposure of NHIS based on district of residence in the survey year. This NHIS exposure variable affects the true NHIS coverage, but should not be correlated with the likelihood of misreporting. Columns 1-3 report results from naive estimators that do not account for misclassification, showing that the standard Probit using self-reported participation (Column 1), the Probit using partially validated data (Column 2), and the restricted sample approach (Column 3), produces dramatically different results in both magnitude and direction. Moreover, the \citet{hausman1998misclassification} method (Column 4), which assumes random misclassification produces, produces results similar to the naive estimators for most variables.\\

\begin{ThreePartTable}
\footnotesize
\begin{TableNotes}[flushleft]
            \footnotesize
            \item \emph{Notes.} Standard errors are clustered at the district level and reported in parentheses. Column 1 reports the MLE estimates using self-reported NHIS participation. Column 2 reports the MLE estimates using the updated (partially validated) NHIS indicator. Column 3 uses same method in Column 2 on a restricted of those whose responses were validated. Column 4 reports results from the \citet{hausman1998misclassification} approach. Column 5 reports the results from the PO MLE approach and Column 6 reports the results from the PPO MLE estimator. Additional variable included in the regressions but excluded from the table for brevity are age, marital status, literacy, rural residence, ethnicity, and number of births in the past 5 years. $*** p \le 0.01$, $** p \le 0.05$, $* p \le 0.10$.
\end{TableNotes}
\begin{longtable}[htp!]{@{}p{4.5cm}P{1.8cm}P{1.8cm}P{2cm}P{1.5cm}P{1.5cm}P{1.8cm}@{}}
\caption{Determinants of NHIS Participation (with Misclassification)}\label{empirical_results} \\ \toprule & \multicolumn{4}{P{7.4cm}}{Naive estimators} & \multicolumn{2}{P{3cm}}{Proposed estimators} \\ \cmidrule{2-5} \cmidrule{6-7} & Probit \newline (Reported) & Probit \newline (Validated) & Probit \newline (Restricted) & HAS & PO MLE & PPO MLE \\ & (1) & (2) & (3) &(4) & (5) & (6) \\ \hline \endfirsthead
\caption[]{Determinants of NHIS Participation (continued)} \\ \toprule & \multicolumn{4}{P{7.4cm}}{Naive estimators} & \multicolumn{2}{P{3cm}}{Proposed estimators} \\ \cmidrule{2-5} \cmidrule{6-7} & Probit \newline (Reported) & Probit \newline (Validated) & Probit \newline (Restricted) & HAS & PO MLE & PPO MLE \\ & (1) & (2) & (3) &(4) & (5) & (6) \\ \hline \endhead
\hline \\ \endfoot
\bottomrule \insertTableNotes \endlastfoot
Years of NHIS Exposure&      -0.083         &      -0.007         &       0.151         &      -0.106         &       0.065         &       0.141         \\
                    &     (0.072)         &     (0.064)         &     (0.126)         &     (0.101)         &     (0.119)         &     (0.098)         \\
1 child             &       0.016         &      -0.076         &      -0.190         &      -0.000         &       0.610         &      -0.154         \\
                    &     (0.088)         &     (0.078)         &     (0.124)         &     (0.116)         &     (0.576)         &     (0.099)         \\
2 children          &      -0.032         &      -0.052         &      -0.088         &      -0.084         &       0.152         &      -0.034         \\
                    &     (0.087)         &     (0.104)         &     (0.161)         &     (0.177)         &     (0.776)         &     (0.117)         \\
3+ children         &      -0.126         &      -0.041         &       0.218         &      -0.165         &      -0.163         &       0.195         \\
                    &     (0.177)         &     (0.163)         &     (0.292)         &     (0.244)         &     (0.750)         &     (0.251)         \\
Middle school graduate&       0.157\sym{*}  &      -0.006         &      -0.318\sym{*}  &       0.164         &      -0.735\sym{*}  &      -0.321\sym{***}\\
                    &     (0.074)         &     (0.086)         &     (0.133)         &     (0.092)         &     (0.375)         &     (0.088)         \\
High school graduate&       0.272\sym{**} &       0.091         &      -0.314         &       0.292\sym{*}  &      -0.529         &      -0.374\sym{**} \\
                    &     (0.103)         &     (0.110)         &     (0.166)         &     (0.141)         &     (0.428)         &     (0.125)         \\
More than high school&       0.292         &       0.150         &      -0.122         &       0.342         &      -0.529         &      -0.321         \\
                    &     (0.159)         &     (0.138)         &     (0.181)         &     (0.216)         &     (0.669)         &     (0.164)         \\
  Catholic          &       0.412\sym{*}  &       0.229         &      -0.556         &       0.515         &      -0.413         &      -0.538\sym{*}  \\
                    &     (0.168)         &     (0.177)         &     (0.321)         &     (0.431)         &     (0.750)         &     (0.246)         \\
  Christian         &       0.274\sym{*}  &       0.024         &      -0.779\sym{*}  &       0.353         &      -0.024         &      -0.635\sym{**} \\
                    &     (0.139)         &     (0.152)         &     (0.311)         &     (0.343)         &     (0.728)         &     (0.230)         \\
  Muslim            &       0.354         &       0.242         &      -0.396         &       0.417         &      -0.587         &      -0.418         \\
                    &     (0.199)         &     (0.192)         &     (0.335)         &     (0.374)         &     (0.760)         &     (0.265)         \\
  Traditional       &       0.030         &      -0.247         &      -0.934\sym{*}  &       0.012         &       1.140         &      -0.567\sym{*}  \\
                    &     (0.171)         &     (0.257)         &     (0.468)         &     (0.269)         &     (1.341)         &     (0.275)         \\
\midrule
Observations        &       3,187         &       3,187         &       1,170         &       3,187         &       3,187         &       3,187         \\
\end{longtable}

\end{ThreePartTable}

We now turn to the two consistent estimators. The PO MLE and the PPO MLE estimators produce qualitatively similar estimates in terms of magnitude and direction, with slight differences in efficiency. For instance, both estimators produce a negative coefficient estimate for the education variables, but those from the PO MLE method are imprecisely estimated. We observe the same pattern of results for the religion variables. In summary, the substantial differences between naive and consistent estimators underscore the importance of misclassification in binary choice models. Our proposed PPO MLE approach and PO MLE methods provide consistent estimates that appear to empirically overcome the misclassification bias from naive approaches. 

\section{Extension to Semiparametric Estimation}\label{semi-parametric}

So far, we have considered binary choice model estimation with misclassified outcome when we assume correct specification of the underlying joint density of the data. However, misclassification and misspecification are different problems, and the latter introduces additional challenges. In this section, we formulate semiparametric estimators for the model that satisfy $\sqrt{n}$-consistency and asymptotic normality as usually desired. The semi-parametric methods offer more flexible alternatives that are less dependent on distributional assumptions, although they also introduce a user-chosen parameter in terms of bandwidth. These estimators are semiparametric in the sense that they make no parametric assumption on the form of the joint distribution generating the disturbances, $(\varepsilon_i,\varepsilon_i^*)$,  in equations (\ref{newmodel}) and (\ref{model2}). We only assume that the choice probability functions for each of the possibly distinguishable alternatives discussed earlier depend on two parametrically specified index functions, $\bm{z}_i^\prime\bm{\beta}_R$ and $\bm{x}_i^\prime\bm{\beta}$.\\

{First, we propose a semiparametric maximum likelihood (SML) version of the PPO MLE discussed earlier and we refer to it as the PPO SML. As discussed earlier, the distinguishable alternatives under \textit{Partial Partial Observability} are a successfully validated participation $(y_i^R=1, y^*_i=1)$,  an unsuccessfully validated participation $(y_i^R=1, y_i^*=0)$,  and an unverified non-participation $(y_i^R=0)$.} The PPO SML is therefore a special case of the general polychotomous choice model of  \citet{lee1995semiparametric}, where the dichotomous indicators for each of our three alternatives are $y_i$, $y_i^R-y_i$, and $1-y_i^R$, and the choice probabilities conditional on the vector of covariates $\bm{w}_i=(\bm{z}_i,\ \bm{x}_i)$ are functions of the vector of indices denoted $\bm{w}_i\bm{\theta}=(\bm{z}_i'\bm{\beta}_R,\ \bm{x}_i'\bm{\beta})$, where $\bm{\theta}=(\bm{\beta}_R,\ \bm{\beta} )$ is the vector of parameters. \\

The semiparametric log-likelihood function to be maximized is defined by:

\begin{equation}
\mathcal{L}_n(\bm{\theta}) = \dfrac{1}{n}\sum_{i=1}^{n}I_{\bm{W}_n}\bigg[ y_i\ln{\bar{P}_{n,i}(\bm{\theta})}+(y_i^R-y_i)\ln{\bar{Q}_{n,i}(\bm{\theta})} +(1-y_i^R)\ln{\bar{R}_{n,i}(\bm{\theta})}\bigg]
\end{equation}
where  the semiparametric probability functions $\bar{P}_{n,i}$, $\bar{Q}_{n,i}$ and $\bar{R}_{n,i}$ are defined by:

\begin{equation}
\begin{aligned}
\bar{P}_{n,i}(\bm{\theta})&= \widehat{\mathbb{E}}[y_i\ |\bm{w}_i\bm{\theta}]=\sum\limits_{\substack{j=1, j\neq i}}^{n}y_j \boldsymbol{H}_{n,j}(\bm{w}_i\bm{\theta}),\\
\bar{Q}_{n,i}(\bm{\theta})&=\widehat{\mathbb{E}}[y_i^R-y_i | \bm{w}_i\bm{\theta}]=\sum\limits_{\substack{j=1, j\neq i}}^{n}(y_j^R-y_j) \boldsymbol{H}_{n,j}(\bm{w}_i\bm{\theta})\\
\bar{R}_{n,i}(\bm{\theta})&=\widehat{\mathbb{E}}[1-y_i^R | \bm{w}_i\bm{\theta}]=\sum\limits_{\substack{j=1, j\neq i}}^{n}(1-y_j^R) \boldsymbol{H}_{n,j}(\bm{w}_i\bm{\theta})
\end{aligned}
\end{equation}
such that $ \boldsymbol{H}_{n,j}(\bm{w}_i\bm{\theta})= \dfrac{K_{h_n}(\bm{w}_j\bm{\theta}-\bm{w}_i\bm{\theta})}{\sum\limits_{j=1, j\neq i}^{n} K_{h_n}(\bm{w}_j\bm{\theta}-\bm{w}_i\bm{\theta})}$, with ${K}_{h_n}(v)=\dfrac{1}{h_n^2}{K}\left(\dfrac{v}{h_n}\right)$, where ${K}(\bm{\cdot})$ is a bivariate kernel function on $\mathbb{R}^2$ and $h_n>0$ is a bandwidth parameter.
The indicator function $I_{\bm{W}_n}(\bm{w}_i)=\mathbb{I}(\bm{w}_i\in \bm{W}_n)$ is defined such that $\bm{W}_n$ is a compact subset of the support of $\bm{w}_i$ that trims the tails of the distribution of $\bm{w}_i$ to make sure that the limit of the denominator in $\boldsymbol{H}_{n,j}(\bm{w}_i\bm{\theta})$ is strictly bounded away from zero. Details about the choice of this indicator are given in the appendix. Notice that $\bar{R}_{n,i}$ can also be defined by $\bar{R}_{n,i}(\bm{\theta})=1- \bar{P}_{n,i}(\bm{\theta})-\bar{Q}_{n,i}(\bm{\theta})$. \\

The semiparametric estimator of $\widehat{\bm{\theta}}=(\widehat{\bm{\beta}}_R, \widehat{\bm{\beta}})$ is therefore defined by: 
\begin{align}\label{sml}
\widehat{\bm{\theta}} &= \argmax_{\bm{\theta}} L_n(\bm{\theta}). 
\end{align}
Under the minimal identification and regularity conditions that we describe in the appendix, 
this estimator is consistent and asymptotically normal. That is,
\[
\sqrt{n}(\widehat{\bm{\theta}}-\bm{\theta}_0) \ \xrightarrow{d}\  N\bigg(0,\, \bm{\Sigma}^{-1}\bm{\Omega}\bm{\Sigma}^{-1}\bigg),
\]
where the expressions for the variance components, $\bm{\Sigma}$ and $\bm{\Omega}$, are given in the appendix.  The asymptotic variance is almost semiparametrically efficient because it is approximately equal to $\bm{\Sigma}^{-1}$, for arbitrary small amounts of trimming.\\

{Second, we also propose a semiparametric version of the PO MLE discussed earlier and we refer to it as the PO SML. Under \textit{Partial Observability}, the two distinguishable alternatives for a respondent are that they participated and their participation was successfully verified  $(y_i=1)$,  or the contrary, that is, they did not participate or their participation was unsuccessfully verified $(y_i=0)$.}\\ 

{
The semiparametric log-likelihood function to be maximized is defined by:
\begin{equation}\label{PO SLE}
\mathcal{L}_n(\bm{\theta}) = \dfrac{1}{n}\sum_{i=1}^{n}I_{\bm{W}_n}\bigg[ y_i\ln{\bar{P}_{n,i}(\bm{\theta})}+(1-y_i)\ln{[1-\bar{P}_{n,i}(\bm{\theta})}]\bigg]
\end{equation}
where all the components in this expression are the same as above. Under the same regularity conditions described above, maximization yields a consistent and asymptotically normal estimator whose variance components are given in the appendix. For the general proof of these asymptotic results, we refer the reader to Appendix A of \citet{lee1995semiparametric} and \citet{ichimura91}. \\
}

\begin{table}[t!]
	\footnotesize
	\centering
	\caption{Normalized coefficients from PPO-SML and PO-SML estimators}
	\label{tab:pposml_posml_norm}
	\begin{threeparttable}		
		\begin{tabular}{@{}P{2cm}P{2cm}P{2cm}P{2.2cm}P{2.2cm}@{}}
\toprule
Average false negative rate & Average false positive rate & True normalized $\beta$ & PPO-SML \newline (normalized coefficient) & PO-SML \newline (normalized coefficient) \\
(1) & (2) & (3) & (4) & (5) \\
\midrule
\multirow{6}{*}{5\%}  & \multirow{3}{*}{5\%}  & 0.947 & 0.858 & 0.851 \\
  &                       & -0.237 & -0.234 & -0.243 \\
  &                       & 0.237 & 0.273 & 0.258 \\
\cmidrule{2-5}
  & \multirow{3}{*}{20\%} & 0.950 & 0.882 & 0.847 \\
  &                       & -0.237 & -0.251 & -0.219 \\
  &                       & 0.237 & 0.230 & 0.193 \\
\cmidrule{1-5}
\multirow{6}{*}{10\%}  & \multirow{3}{*}{5\%}  & 0.983 & 0.872 & 0.935 \\
  &                       & -0.246 & -0.233 & -0.257 \\
  &                       & 0.246 & 0.187 & 0.303 \\
\cmidrule{2-5}
  & \multirow{3}{*}{20\%} & 0.960 & 0.830 & 0.853 \\
  &                       & -0.240 & -0.261 & -0.264 \\
  &                       & 0.240 & 0.334 & 0.241 \\
\cmidrule{1-5}
\multirow{6}{*}{20\%}  & \multirow{3}{*}{5\%}  & 0.915 & 0.846 & 0.803 \\
  &                       & -0.229 & -0.260 & -0.248 \\
  &                       & 0.229 & 0.335 & 0.222 \\
\cmidrule{2-5}
  & \multirow{3}{*}{20\%} & 0.959 & 0.860 & 0.846 \\
  &                       & -0.240 & -0.242 & -0.242 \\
  &                       & 0.240 & 0.292 & 0.266 \\
\bottomrule
\end{tabular}

		\begin{tablenotes}[flushleft]
			\footnotesize
			\item \emph{Notes.} Table compares the semiparametric maximum likelihood estimates for the PO and PPO discussed in Section \ref{semi-parametric}. Columns 1 and 2 report average misclassification rates. 
            Each cell reports the normalized coefficient estimates from the two SML estimators. The last element of the coefficient vector is normalized to 1 in the semiparametric estimation, and thus, omitted from the table. 
		\end{tablenotes}
	\end{threeparttable}
\end{table}

Table~\ref{tab:pposml_posml_norm} reports simulation results for the above
semiparametric maximum likelihood estimators.
The simulation design is identical to the Monte Carlo experiment reported in Table \ref{tab1}, where we presented parametric estimates. In particular, we use the same values for the true coefficients
$(\beta_1,\beta_2,\beta_3,\beta_4) = (2,-0.5,0.5,1)$, with the only difference being that Table~\ref{tab:pposml_posml_norm} focuses on the PPO-SML and PO-SML estimators. Both SML estimators are implemented using a bivariate Gaussian kernel, with the conditional probabilities estimated via Nadaraya–Watson-type kernel averages. For simplicity, we use a bandwidth $h$ of the form $h = cN^{-p},$ with $c=1$ and $p=6.5.$ \\

Following \citet{abrevaya1999semiparametric}, all coefficients reported in Table~\ref{tab:pposml_posml_norm} are \emph{normalized} to allow
for a meaningful comparison of estimates across methods. Specifically, the normalization ensures that the PPO-SML estimate vector has a unit length, that is, $\hat\beta_{1}^{2} + \hat\beta_{2}^{2} + \hat\beta_{3}^{2} + \hat\beta_{4}^{2}=1$. We then apply this normalization to the true coefficient vector and the PO-SML estimates, ensuring that the simulation design and the estimators are compared on the same scale. In Table~\ref{tab:pposml_posml_norm}, we can see that PPO-SML and PO-SML deliver broadly similar performance across designs, producing estimates reasonably close to the true normalized coefficients. Even in the highest misclassification cases, the differences between PPO-SML and PO-SML are modest. In sum, both semiparametric estimators are promising alternatives for model misspecification when accounting for misclassification in binary choice models.

\section{Conclusion}\label{conclusion}

This paper addresses the problem of estimating binary choice models when the dependent variable is subject to endogenous misclassification that can be partially validated. We formalize the concept of partial validation, where survey features allow  researchers to verify responses for only one category of a binary outcome, and demonstrate how this transforms a complex two-sided misclassification into a more tractable one-sided problem. Our theoretical analysis shows that partial validation alone does not solve or mitigate misclassification bias, and using partially validated measures in standard binary choice estimation can still lead to sign reversals in estimated effects. To address this challenge, we propose two Maximum Likelihood Estimators under partial observability that exploit the structure created by partial validation to consistently estimate model parameters when misclassification errors are allowed to depend on covariates and to be correlated with true responses. \\ 

Monte Carlo simulation results provide compelling evidence for our approach. In particular, we find that endogenous misreporting where misclassification probabilities are covariate-dependent creates severe estimation challenges for conventional approaches. In contrast, our proposed estimators perform well for various degrees of the severity of misclassification. We provide an empirical illustration using national health insurance participation in Ghana which was misreported by respondents in the 2014 Demographic and Health Survey.  Relative to the proposed consistent estimators, we find that naive estimators provide a misleading picture of the determinants of health insurance take-up. Meanwhile, both the proposed PO MLE and PPO MLE estimators provide comparable signs and magnitudes, although coefficient significance may slightly differ across methods. We develop semiparametric versions of our proposed parametric methods that relax the distributional assumptions to overcome potential model misspecification. Simulations show that these methods perform satisfactorily well and provide useful alternatives for simultaneously addressing misclassification and model misspecification in binary choice models.\\

Our paper has important implications for survey design and empirical practice. Our results suggest that incorporating verification questions for even one response category can enable researchers to obtain consistent estimates of binary choice models and other regression models in the presence of endogenous misclassification. Future work could explore how the proposed method can be applied more broadly, particularly in the case of ordered or multinomial choice models with misclassified outcomes. 

\clearpage
\begin{singlespace}
	\bibliography{biblio}
\end{singlespace}

\newpage
\begin{appendices}

\section{Mathematical Proofs}

\subsection{Proof of Lemma \ref{A1}}
\begin{proof}
 The indicator of truthfulness for those who report program participation can be written as
\begin{equation}\label{truth}
t_i  = \begin{cases}
          y_i^*  y^R_i & if ~y^R_i=1 \\
         unobserved & if ~ y^R_i=0
         \end{cases}
\end{equation}
This implies, by Equation \eqref{PV}, that  

\begin{equation*}
y_i  = 
\begin{cases}
y_i^*  y^R_i & if ~ y^R=1  \\
 0 & if ~ y_i^R=0\\
\end{cases}
\end{equation*}

That is,
\begin{equation}\label{PV2}
y_i= y_i^*  y^R_i,
\end{equation}

which, from equation \eqref{yR}, implies 
\[
y_i= y_i^*\left\{y^*_{i}(1 - d_{i}) + (1 -  y^*_{i})d_{i}\right\}= y_i^* (1-d_i)
\]
\end{proof}

\subsection{Proof of Theorem \ref{theoremR}}
\begin{proof}
The first order conditions for maximizing the likelihood function $\mathcal{L}_n^R$ is given by
\begin{align*}
\frac{\partial \mathcal{L}_n^R(\hat{\bm{\beta}}_R)}{\partial {\bm{\beta}}} = S_n(\bm{\hat{\beta}}_R)
& = \sum_{i=1}^n \left[\left(\frac{y^R_i - F(\bm{x_i\hat{\beta}}_R)}{F(\bm{x_i\hat{\beta}}_R)(1-F(\bm{x_i\hat{\beta}}_R))}  \right)f(\bm{x_i\hat{\beta}}_R) \bm{x_i} \right] = 0
\end{align*}

By a first order Taylor expansion of $S_n(\hat{\bm{\beta}}_R)$ around the true parameter vector, $\bm{\beta}_0$, we can write 
\begin{align}\label{taylor}
0 = S_n(\hat{\bm{\beta}}_R) = S_n(\bm{\beta}_0) + \frac{\partial S_n(\tilde{\bm{\beta}}_R)}{\partial \bm{\beta}}\left( \hat{\bm{\beta}}_R - \bm{\beta}_0\right), 
\end{align}
where $\tilde{\bm{\beta}}_R$ is a convex combination of $\hat{\bm{\beta}}_R$ and $\bm{\beta}_0$. The continuous differentiability and concavity of the likelihood function guarantees that $\dfrac{\partial S_n(\tilde{\bm{\beta}}_R)}{\partial \bm{\beta}}$ is negative definite and $-\dfrac{1}{n}\dfrac{\partial S(\tilde{\bm{\beta}}_R)}{\partial \bm{\beta}}$ converges to  $H(\tilde{\bm{\beta}})$, the Hessian evaluated at $\tilde{\bm{\beta}}$ which is the probability limit of $\tilde{\bm{\beta}}_R$ (i.e., $\tilde{\bm{\beta}}$ is a convex combination of $\bm{\beta}_R$ and $\bm{\beta}_0$.)

From equation \eqref{taylor}, we have 
\begin{align*}
\hat{\bm{\beta}}_R  - \bm{\beta}_0 &= \left[-\frac{1}{n}\frac{\partial S_n(\bar{\bm{\beta}}_R)}{\partial \bm{\beta}}\right]^{-1}\frac{S_n(\bm{\beta}_0)}{n} \overset{a}{ = } \left[H(\tilde{\bm{\beta}})\right]^{-1}\frac{S_n(\bm{\beta}_0)}{n}, 
\end{align*}
where $\overset{a}{ = }$ denotes asymptotic equivalence. The asymptotic bias in $\hat{\bm{\beta}}_R$ is then characterized by evaluating the probability limit of the quantity $\dfrac{S_n(\bm{\beta}_0)}{n}, $ the average score (gradient) vector, where we note that

\begin{align}\label{avgscore}
\dfrac{S_n(\bm{\beta}_0)}{n}
=\left[\left(\frac{y^R_i - F(\bm{x_i{\beta}}_0)}{F(\bm{x_i{\beta}}_0)(1-F(\bm{x_i{\beta}}_0))}  \right)f(\bm{x_i{\beta}}_0) \bm{x_i} \right] 
\end{align}

By the law of large numbers, Equation \eqref{avgscore} converges in probability limit to:
\begin{align*}
\esp (s_i(\bm{\beta})) &= \esp \left[\left\{\frac{y_i^R - F(\bm{x_i\beta}_0)}{F(\bm{x_i\beta}_0)(1-F(\bm{x_i\beta}_0))}  \right\}f(\bm{x_i\beta}_0) \bm{x_i} \right] \\ 
&= \esp \left[\left\{\frac{\esp[y_i^R|\bm{z}_i] - F(\bm{x_i\beta}_0)}{F(\bm{x_i\beta}_0)(1-F(\bm{x_i\beta}_0))}  \right\}f(\bm{x_i\beta}_0) \bm{x_i} \right] (\text{by iterated expectations}) 
\end{align*}
Now, notice that 
\begin{equation}\label{EYR}
\esp[y_i^R|\bm{z}_i]=a_0^R(\bm{z}_i)+\left(1-a_0^R(\bm{z}_i)-a_1^R(\bm{z}_i)\right)F(\bm{x_i\beta}_0)
\end{equation}
Hence, 
\begin{align*}
\esp (s_i(\bm{\beta})) &= \esp \left[\left\{\frac{a_0^R(\bm{z}_i)+\left(1-a_0^R(\bm{z}_i)-a_1^R(\bm{z}_i)\right)F(\bm{x_i\beta}_0) - F(\bm{x_i\beta}_0)}{F(\bm{x_i\beta}_0)(1-F(\bm{x_i\beta}_0))}  \right\}f(\bm{x_i\beta}_0) \bm{x_i} \right]\\
&=-\esp\left[\left\{\frac{a_1^R(\bm{z}_i)}{1 - F(\bm{x}_i'\bm{\beta}_0)}-\frac{a_0^R(\bm{z}_i)}{ F(\bm{x}_i'\bm{\beta}_0)}\right\}f(\bm{x}_i'\bm{\beta}_0) \bm{x_i} \right]
\end{align*}

As for the hessian, it is obtained from taking the probability limit of the second order log-likelihood derivative, $-\dfrac{1}{n}\dfrac{\partial S_n(\tilde{\bm{\beta}}_R)}{\partial \bm{\beta}}$,  evaluated at  $\tilde{\bm{\beta}}$. This derivative is given by the main element:

\[
\dfrac{\partial s_i(\tilde{\bm{\beta}}_R)}{\partial \bm{\beta}}=f(\bm{x}_i'\tilde{\bm{\beta}}_R) \bm{x_i}\dfrac{\partial U_i(\tilde{\bm{\beta}}_R)}{\partial \bm{\beta}'}+U_i(\tilde{\bm{\beta}}_R)\dfrac{\partial f(\bm{x}_i'\tilde{\bm{\beta}}_R)}{\partial \bm{\beta}}\bm{x}_i',
\]
where
\[
U_i(\tilde{\bm{\beta}}_R)=\frac{y^R_i - F(\bm{x_i\tilde{\beta}}_R)}{F(\bm{x_i\tilde{\beta}}_R)(1-F(\bm{x_i\tilde{\beta}}_R))},
\] 
\[
\dfrac{\partial U_i(\tilde{\bm{\beta}}_R)}{\partial \bm{\beta}'}=-\frac{F(\bm{x_i\tilde{\beta}}_R)\left(1-F(\bm{x_i\tilde{\beta}}_R)\right)+\left(1-2F(\bm{x_i\tilde{\beta}}_R\right)\left(y^R_i - F(\bm{x_i\tilde{\beta}}_R)\right)}{F(\bm{x_i\tilde{\beta}}_R)^2(1-F(\bm{x_i\tilde{\beta}}_R))^2}f(\bm{x}_i'\tilde{\bm{\beta}}_R)\bm{x}_i',
\]
and
\[
\dfrac{\partial f(\bm{x}_i'\tilde{\bm{\beta}}_R)}{\partial \bm{\beta}}=\dfrac{d f(\bm{x}_i'{\bm{\beta}})}{d (\bm{x}_i'{\bm{\beta}})}\bigg|_{{\bm{\beta}}=\tilde{\bm{\beta}}_R}\bm{x}_i=f^{(1)}(\bm{x}_i'\tilde{\bm{\beta}}_R)\bm{x}_i
\]
The Hessian is then

\begin{align*}
\bm{A}_R=H(\tilde{\bm{\beta}})&=\pl~ \left\{-\dfrac{1}{n}\dfrac{\partial S_n(\tilde{\bm{\beta}}_R)}{\partial \bm{\beta}}\right\}=-\esp\left[\dfrac{\partial s_i(\tilde{\bm{\beta}})}{\partial \bm{\beta}}\right]\\
&=-\esp\left[f(\bm{x}_i'\tilde{\bm{\beta}}) \bm{x_i}\dfrac{\partial U_i(\tilde{\bm{\beta}})}{\partial \bm{\beta}'}+U_i(\tilde{\bm{\beta}})\dfrac{\partial f(\bm{x}_i'\tilde{\bm{\beta}})}{\partial \bm{\beta}}\bm{x}_i',  \right]\\
&=\esp\left[f(\bm{x}_i'\tilde{\bm{\beta}}) \bm{x_i}\esp\left[-\dfrac{\partial U_i(\tilde{\bm{\beta}})}{\partial \bm{\beta}'}\bigg|\bm{z}_i\right]-\esp[U_i(\tilde{\bm{\beta}})|\bm{z}_i]\dfrac{\partial f(\bm{x}_i'\tilde{\bm{\beta}})}{\partial \bm{\beta}}\bm{x}_i',  \right]
\end{align*}

where, by iterated expectations and Equation \eqref{EYR}, we have 
\[
-\esp[U_i(\tilde{\bm{\beta}})|\bm{z}_i]=-\frac{\esp[y^R_i|\bm{z}_i] - F(\bm{x_i\tilde{\beta}})}{F(\bm{x_i\tilde{\beta}})(1-F(\bm{x_i\tilde{\beta}}))}
=\frac{a_1^R(\bm{z}_i)}{1 - F(\bm{x_i'\tilde{\beta}})}-\frac{a_0^R(\bm{z}_i)}{ F(\bm{x_i'\tilde{\beta}})},
\]

\begin{align*}
\esp\left[-\frac{\partial U_i(\tilde{\bm{\beta}})}{\partial \bm{\beta}'}\bigg|\bm{z}_i\right]&=\frac{F(\bm{x_i\tilde{\beta}})(1-F(\bm{x_i\tilde{\beta}}))+(1-2F(\bm{x_i\tilde{\beta}})(\esp[y^R_i|\bm{z}_i] - F(\bm{x_i\tilde{\beta}}))}{F(\bm{x_i\tilde{\beta}})^2(1-F(\bm{x_i\tilde{\beta}}))^2}f(\bm{x}_i'\tilde{\bm{\beta}})\bm{x}_i',\\
&= \frac{\tilde{F}_i(1-\tilde{F}_i)+(1-2\tilde{F}_i)\left[a_0^R(\bm{z}_i)(1-\tilde{F}_i)-a_1^R(\bm{z}_i)\tilde{F}_i\right]}{\tilde{F}_i^2(1-\tilde{F}_i)^2}f(\bm{x}_i'\tilde{\bm{\beta}})\bm{x}_i',
\end{align*}
 and $\tilde{F}_i=F(\bm{x_i'\tilde{\beta}})$. 

\end{proof}


\subsection{Proof of Theorem \ref{theoremP}}
\begin{proof}
The proof of this result follows exactly the same argument as for Theorem \ref{theoremR}, except that here we are dealing with a one-sided misreporting problem. Indeed, notice that we can write:
\[
\esp[y_i|\bm{z}_i]=\left(1-a_1(\bm{z}_i)\right)F(\bm{x_i'{\beta}}_0))
\]
which is similar to Equation \eqref{EYR} except that $a_0(\bm{z}_i)=0$ here. Hence, the results for Theorem \ref{theoremP} can be obtained by replacing $a_1^R(\bm{z}_i)$ by $a_1(\bm{z}_i)$, $a_0(\bm{z}_i)$ by $0$, and $\tilde{\bm{\beta}}$ by $\bar{\bm{\beta}}$ in all the steps used to derive Theorem \ref{theoremR}.  
\end{proof}

\bigskip


\section{Additional details on the semiparametrics}

\subsection{Trimming Process}
For the trimming function $I_{n}(\bm{w}_i)$,  \citet{lee1995semiparametric} recommends to define the trimming  set $\bm{W_n}$ as $\bm{W_n}=\{\bm{s}_i: \xi_{np}\leq \bm{s}_i \leq \xi_{n(1-p)}\}$, where $\bm{s}_i$ is the subvector of all continuous regressors in $\bm{w}_i$, and $\xi_{np}$ is the $p^{th}$ sample quantile vector of $\bm{s},\ 0<p<1$. This basically means restricting continuous observations between their $p^{th}$ and $(1-p)^{th}$ percentiles. Note that, the set $\bm{W}_n$ converges to $\bm{W}=\{\bm{s}_i:\ \xi_{p}\leq \bm{s}_i \leq \xi_{(1-p)}\}$ in probability; and $\bm{W}$ is a proper compact subset of the support of $\bm{w}_i$. While some sample information will be lost by the trimming, the loss can be small as the trimmed quantile can be choosen to be an arbitrarily small positive number. 

\subsection{Numerical Algrorithm}

A practical numerical challenge that may sometimes arise in the estimation of the SML estimator of $\hat{\bm{\theta}}$ is that the Kernel-based estimates of  $\bar{P}_{n,i}(\bm{\theta})$, $\bar{Q}_{n,i}(\bm{\theta})$ and $\bar{Q}_{n,i}(\bm{\theta})$  are not necessarily positive and less than 1. One way to overcome this issue is to implement a procedure that systematically modifies the objective function for those ill-behaved situations. A possible modification of the likelihood function above is therefore:
\begin{align*}
\mathcal{L}_n(\bm{\theta}) = \frac{1}{n}\sum_{i=1}^{n}I_{n}(\bm{w_i})\bigg[ y_i\ln{\bar{P}_{n,i}(\bm{\theta})} \ind(\bigtriangleup_n<\bar{P}_{n,i}(\bm{\theta}) <1- \bigtriangleup_n) \\+ (y_i^R-y_i)\ln{\bar{Q}_{n,i}(\bm{\theta})}\ind(\bigtriangleup_n<\bar{Q}_{n,i}(\bm{\theta}) <1- \bigtriangleup_n)  \\
+(1-y_i^R)\ln{\bar{R}_{n,i}(\bm{\theta})}\ind(\bar{R}_{n,i}(\bm{\theta})  <1- \bigtriangleup_n)\bigg]
\end{align*}

where $\bigtriangleup_n$ is a specified deterministic sequence such that $\lim_{n \to \infty} \bigtriangleup_n=0$.

\subsection{Asymptotic distribution}
	Assume the regularity conditions are satisfied.\\
    
For the PPO SML, we have 
	$\sqrt{n}(\widehat{\bm{\theta}}-\bm{\theta}_0) \ \xrightarrow{d}\  N\bigg(0, \Sigma^{-1}\Omega\Sigma^{-1}\bigg), $
	where
	\begin{align*}
	\Sigma &= \mathbb{E}\bigg[I_W\frac{1}{\bar{P}_i}\frac{\partial{\bar{P}_i(\bm{\theta}_0)}}{\partial{\bm{\theta}}} \frac{\partial{\bar{P}_i(\bm{\theta}_0)}}{\partial{\bm{\theta}^\prime}} + I_W\frac{1}{\bar{Q}_i}\frac{\partial{\bar{Q}_i(\bm{\theta}_0)}}{\partial{\bm{\theta}}} \frac{\partial{\bar{Q}_i(\bm{\theta}_0)}}{\partial{\bm{\theta}^\prime}}+I_W\frac{1}{\bar{R}_i}\frac{\partial{\bar{R}_i(\bm{\theta}_0)}}{\partial{\bm{\theta}}} \frac{\partial{\bar{R}_i(\bm{\theta}_0)}}{\partial{\bm{\theta}^\prime}}\bigg],\\
\text{and}\\
	\Omega &= \mathbb{E} \bigg\{\frac{1}{\bar{P}_i}\bigg[I_W\frac{\partial{\bar{P}_i}}{\partial{\bm{\theta}}}\frac{\partial{\bar{P}_i}}{\partial{\bm{\theta}^\prime}}-\mathbb{E}\bigg(I_W\frac{\partial{\bar{P}_i}}{\partial{\bm{\theta}}}\ \bigg|\ \bm{w}_i\bm{\theta}\bigg)\mathbb{E}\bigg(I_W\frac{\partial{\bar{P}_i}}{\partial{\bm{\theta}^\prime}}\ \bigg|\ \bm{w}_i\bm{\theta}\bigg)\bigg]\bigg\}\\
 &\quad \quad +   \mathbb{E} \bigg\{\frac{1}{\bar{Q}_i}\bigg[I_W\frac{\partial{\bar{Q}_i}}{\partial{\bm{\theta}}}\frac{\partial{\bar{Q}_i}}{\partial{\bm{\theta}^\prime}}-\mathbb{E}\bigg(I_W\frac{\partial{\bar{Q}_i}}{\partial{\bm{\theta}}}\ \bigg|\ \bm{w}_i\bm{\theta}\bigg)\mathbb{E}\bigg(I_W\frac{\partial{\bar{Q}_i}}{\partial{\bm{\theta}^\prime}}\ \bigg|\ \bm{w}_i\bm{\theta}\bigg)\bigg]\bigg\}\\
 &\quad \quad \quad \quad +  \mathbb{E} \bigg\{\frac{1}{\bar{R}_i}\bigg[I_W\frac{\partial{\bar{R}_i}}{\partial{\bm{\theta}}}\frac{\partial{\bar{R}_i}}{\partial{\bm{\theta}^\prime}}-\mathbb{E}\bigg(I_W\frac{\partial{\bar{R}_i}}{\partial{\bm{\theta}}}\ \bigg|\ \bm{w}_i\bm{\theta}\bigg)\mathbb{E}\bigg(I_W\frac{\partial{\bar{R}_i}}{\partial{\bm{\theta}^\prime}}\ \bigg|\ \bm{w}_i\bm{\theta}\bigg)\bigg]\bigg\}
	\end{align*}

If $I_W=1,$ then 
	\begin{align*}
	\Omega &= \mathbb{E}\bigg[\frac{1}{\bar{P}_i}\frac{\partial{\bar{P}_i}}{\partial{\bm{\theta}}} \frac{\partial{\bar{P}_i}}{\partial{\bm{\theta}^\prime}} + \frac{1}{\bar{Q}_i}\frac{\partial{\bar{Q}_i}}{\partial{\bm{\theta}}} \frac{\partial{\bar{Q}_i}}{\partial{\bm{\theta}^\prime}}+\frac{1}{\bar{R}_i}\frac{\partial{\bar{R}_i}}{\partial{\bm{\theta}}} \frac{\partial{\bar{R}_i}}{\partial{\bm{\theta}^\prime}}\bigg]=\Sigma
	\end{align*}
	
Hence the asymptotic variance of $\widehat{\bm{\theta}}$ is $\Sigma^{-1}$.\\
	
This means that for our estimation method, if $I_W(w_i)$ were identically equal to 1 for all $\bm{w}_i$, then our SML, $\widehat{\bm{\theta}}$, would attain the semiparametric efficiency bound in the form discussed by \citet{chamberlain1987asymptotic,thompson1993some},  and \citet{klein1993efficient}. However, in our case, sample information was lost through trimming, leading to a less efficient estimator. The asymptotic effect of trimming drifts the variance expression through the terms $\mathbb{E}\left(I_W \dfrac{\partial{\bar{P}_i}}{\partial{\bm{\theta}}} |\ \bm{w}_i\bm{\theta}\right)$, $\mathbb{E}\left(I_W \dfrac{\partial{\bar{Q}_i}}{\partial{\bm{\theta}}} |\ \bm{w}_i\bm{\theta}\right)$ and $\mathbb{E}\left(I_W \dfrac{\partial{\bar{Q}_i}}{\partial{\bm{\theta}}} |\ \bm{w}_i\bm{\theta}\right)$. Since these terms shrink towards zero as the trimming quantile $p$ gets smaller, the loss of efficiency can be small when an arbitrarily small percentage of observations are trimmed.\footnote{Given that the function $\Gamma_p \xrightarrow {p}\Sigma^{-1}_p\Omega_p\Sigma^{-1}_p$ is continuous such that $\lim_{p\to 0}\Gamma_p=\Sigma^{-1}$, then $\Sigma^{-1}$ can well approximate the variance of $\hat{\theta}$ when $p$ is small.} A consistent estimator of the variance is obtained using the sample counterparts $\widehat{\Omega}$ and $\widehat{\Sigma},$ where $\bm{\theta}_0$ is replaced by $\widehat{\bm{\theta}}$ and 
\[
\dfrac{\partial{{\bar{P}}}_{n,i}}{\partial{\bm{\theta}}}= \sum_{j,j\neq i}y_j\dfrac{\partial \bm{H}_{n,j}}{\partial{\bm{\theta}}} ,\quad 	\dfrac{\partial{{\bar{Q}}}_{n,i}}{\partial{\bm{\theta}}}=\sum_{j,j\neq i}(y_j^R-y_j)\dfrac{\partial \bm{H}_{n,j}}{\partial{\bm{\theta}}},\quad 	\dfrac{\partial{{\bar{R}}}_{n,i}}{\partial{\bm{\theta}}}= \sum_{j,j\neq i}(1-y^R_j)\dfrac{\partial \bm{H}_{n,j}}{\partial{\bm{\theta}}}
\]
	including all the trimming $I_{\bm{W}_n}$ and $\xi_n$.\\


For the PO SML, we can also write $\sqrt{n}(\widehat{\bm{\theta}}-\bm{\theta}_0) \ \xrightarrow{d}\  N\bigg(0, \Sigma^{-1}\Omega\Sigma^{-1}\bigg), $
	where the components of the asymptotic variance are given by:
   	\begin{align*}
	\Sigma &= \mathbb{E}\bigg[I_W\frac{1}{\bar{P}_i(1-\bar{P}_i)}\frac{\partial{\bar{P}_i(\bm{\theta}_0)}}{\partial{\bm{\theta}}} \frac{\partial{\bar{P}_i(\bm{\theta}_0)}}{\partial{\bm{\theta}^\prime}}\bigg],\\
\text{and}\\
	\Omega &= \mathbb{E} \bigg\{\frac{1}{\bar{P}_i(1-\bar{P}_i)}\bigg[I_W\frac{\partial{\bar{P}_i}}{\partial{\bm{\theta}}}\frac{\partial{\bar{P}_i}}{\partial{\bm{\theta}^\prime}}-\mathbb{E}\bigg(I_W\frac{\partial{\bar{P}_i}}{\partial{\bm{\theta}}}\ \bigg|\ \bm{w}_i\bm{\theta}\bigg)\mathbb{E}\bigg(I_W\frac{\partial{\bar{P}_i}}{\partial{\bm{\theta}^\prime}}\ \bigg|\ \bm{w}_i\bm{\theta}\bigg)\bigg]\bigg\}.\\
    \end{align*}
Likewise, if $I_W=1$, then 
\[
\Sigma = \mathbb{E}\bigg[\frac{1}{\bar{P}_i(1-\bar{P}_i)}\frac{\partial{\bar{P}_i(\bm{\theta}_0)}}{\partial{\bm{\theta}}} \frac{\partial{\bar{P}_i(\bm{\theta}_0)}}{\partial{\bm{\theta}^\prime}}\bigg]=\Omega,
\] so that the asymptotic variance of $\widehat{\bm{\theta}}$ is $\Sigma^{-1}$.
\end{appendices}
\end{document}